\documentclass[12pt]{iopart}
\usepackage{graphicx} 
\usepackage{dcolumn} 
\usepackage{amssymb}
\usepackage{iopams}

\begin{document}

\title[Schwarzschild Black Hole Threaded by a Cosmic String]{Renormalized Vacuum Polarization and Stress Tensor on the Horizon of a Schwarzschild Black Hole Threaded by a Cosmic String}
\author{Adrian C. Ottewill and Peter Taylor}
\address{School of Mathematical Sciences \& Complex and Adaptive Systems Laboratory, University College Dublin, Belfield, Dublin 4, Ireland}
\eads{\mailto{adrian.ottewill@ucd.ie} and \mailto{peter.taylor@ucd.ie}}
\date{\today}
\begin{abstract}
We calculate the renormalized vacuum polarization and stress tensor for a massless, arbitrarily coupled scalar field in the Hartle-Hawking vacuum state on the horizon of a Schwarzschild black hole threaded by an infinte straight cosmic string. This calculation relies on a generalized Heine identity for non-integer Legendre functions which we derive without using specific properties of the Legendre functions themselves. \end{abstract}
\maketitle


\section{Introduction}
There is a long and successful history of  calculating the renormalized vacuum polarization and stress-energy tensor on spherically-symmetric black holes (see \cite{Candelas:1980zt,Candelas:1984pg,Anderson:1989vg,JensenOttewill, JMO1, JMO2, Anderson:1990jh,Winstanley:2007}), however, in the astrophysically significant case of the Kerr-Newman black hole, such calculations  have proved intractable. This is due to the fact that all the calculations in the spherically symmetric case have relied heavily on specific properties of the Legendre functions (mainly the Legendre Addition Theorem and the Heine Identity) and also the applicability of the Watson-Sommerfeld Formula to obtain a mode-sum expression that is numerically tractable. In the axially symmetric case, however, addition theorems for the angular functions do not exist nor do we have an analog of the Heine identity for the angular functions that arise. We will show, in the framework of the Schwarzschild black hole threaded by a cosmic string, that we can proceed in the axially symmetric case by obtaining useful summation formulae based on the Hadamard structure of the Green function without using specific properties of the mode-functions themselves. In this paper we will adopt this approach in order to derive a generalization of the standard Heine identity, adapted to the non-integer Legendre functions which arise in the separation of the wave equation on the background metric of a Schwarzschild black hole threaded by a cosmic string. Furthermore, we apply this formula to the calculation of the vacuum polarization and renormalized stress tensor for a massless scalar field on the horizon of the black hole.


\section{The Green's Function}
 The cosmic string is modeled by introducing an azimuthal deficit parameter, $\alpha$, into the standard Schwarzschild metric. We
may describe this in  coordinates $(t, r, \theta, \tilde{\phi})$, where  $\tilde{\phi}$ is periodic with period  $2\pi \alpha$, so we may take $\tilde{\phi}\in[0,  2\pi \alpha)$, in which the line element is given by
\begin{equation}
\fl
{\rmd}s^2 = -(1-2M/r) {\rmd}t^2 +(1-2M/r)^{-1} {\rmd}r^2 +r^2 {\rmd} \theta^2 +  r^{2} \sin^{2}\theta {\rmd} \tilde{\phi}^{2}.
\end{equation}
For GUT scale cosmic strings $\alpha = 1- 4 \mu$ where $\mu$ is the mass per unit length of the string so we shall assume $0 < \alpha \le 1$.
We can alternatively define a new azimuthal coordinate by
\begin{equation}
\tilde{\phi} = \alpha \phi
\end{equation}
so that  $\phi$ is periodic with period  $2\pi$ and we may take $\phi \in [0, 2 \pi)$. In coordinates $(t, r, \theta, \phi)$,  the line element is given by
\begin{equation}
\label{eq:metric}
\fl
{\rmd}s^2 = -(1-2M/r) {\rmd}t^2 +(1-2M/r)^{-1} {\rmd}r^2 +r^2 {\rmd} \theta^2 + \alpha^{2} r^{2} \sin^{2}\theta {\rmd} \phi^{2}.
\end{equation}

We shall consider a massless, minimally coupled scalar field, $\varphi$, in the Hartle-Hawking vacuum state. 
Since this is a thermal state, it is convenient to work with the Euclidean Green's function, performing a Wick rotation of the
 temporal coordinate $t\rightarrow -i \tau$ and eliminating the conical singularity at $r=2M$ by making $\tau$ periodic with period
$2\pi/\kappa$ where $\kappa=1/(4M)$ is  the surface gravity of the black hole. 
The massless, minimally coupled scalar field, satisfies the homogenous wave-equation
\begin{equation}
\Box \varphi(\tau,r,\theta,\phi)=0,
\end{equation}
which can be solved by a separation of variables by writing
\begin{equation}
\varphi(\tau,r,\theta,\phi)\sim \rme^{i n \kappa \tau+i m\phi} P(\theta)R(r)
\end{equation}
where $P(\theta)$ is regular and satisfies 
\begin{equation}
\label{eq:legendre}
\Big\{\frac{1}{\sin\theta} \frac{{\rmd}}{{\rmd} \theta}\Big(\sin\theta\frac{{\rmd}}{{\rmd} \theta}\Big) -\frac{m^{2}}{\alpha^{2}\sin^{2}\theta}+\lambda(\lambda+1)\Big\}P(\theta)=0
\end{equation}
while $R(r)$ satisfies
\begin{equation}
\label{eq:radialhomogeneous}
\Big\{\frac{{\rmd}}{{\rmd} r}(r^2-2Mr)\frac{{\rmd}}{{\rmd}r} - \lambda(\lambda+1)-\frac{n^2 \kappa^2 r^4}{r^2-2Mr}\Big\}R(r)= 0.
\end{equation}
 The $\lambda(\lambda+1)$ term arises as the separation constant. The choice of $\lambda$ is arbitrary for $\varphi$ to satisfy the wave equation but requires a specific choice in order for the mode-function $P(\theta)$ to satisfy the boundary conditions of regularity on the poles. In the Schwarzschild case (no cosmic string, $\alpha=1$), regularity on the poles means that $\lambda=l$, i.e the separation constant is $l(l+1)$. In the cosmic string case, the appropriate choice of $\lambda$ that guarantees regularity of the angular functions on the poles is
\begin{equation}
\label{eq:lambda}
\lambda=l-|m|+|m|/\alpha.
\end{equation}  
(We note here the dependence of $\lambda$ on $l$ and $m$. However, to avoid being typographically cumbersome, we will leave it implicit in the remainder of this paper.) 
With this choice of $\lambda$, the angular function is the Legendre function of both non-integer order and non-integer degree,\textit{viz.},
\begin{equation}
P(\theta)=P_{\lambda}^{-|m|/\alpha}(\cos\theta).
\end{equation}
We have shown in the Appendix that these angular functions satisfy the following normalization condition,
\begin{eqnarray}
\label{eq:norm}
&\int_{-1}^{1}P_{l-|m|+|m|/\alpha}^{-|m|/\alpha}(\cos\theta) P_{l'-|m|+|m|/\alpha}^{-|m|/\alpha}(\cos\theta) d(\cos\theta) 
=\nonumber\\
&\qquad\qquad\qquad\frac{2}{(2\lambda+1)}\frac{\Gamma(\lambda-|m|/\alpha+1)}{\Gamma(\lambda+|m|/\alpha+1)}\delta_{ll'}. 
\end{eqnarray}

The periodicity of the Green's function with respect to $(\tau-\tau')$ and $(\phi-\phi')$ with periodicity $2\pi/\kappa$ and $2\pi$, respectively, combined with Eq.(\ref{eq:norm}) allow us to write the mode-sum expression for the Green's function as
\begin{eqnarray}
\label{eq:greensfn}
\fl
G(x,x')=\frac{\kappa}{8\pi^{2}}\sum_{n=-\infty}^{\infty} {\rme}^{i n \kappa (\tau-\tau')}&\sum_{m=-\infty}^{\infty} {\rme}^{im(\phi-\phi')} \sum_{l=|m|}^{\infty} (2\lambda+1)\frac{\Gamma(\lambda+|m|/\alpha+1)}{\Gamma(\lambda-|m|/\alpha+1)}  \nonumber\\
 &\qquad P_{\lambda}^{-|m|/\alpha}(\cos\theta) P_{\lambda}^{-|m|/\alpha}(\cos\theta')\chi_{n\lambda}(r,r'),
\end{eqnarray}
where $\chi_{n\lambda}(r,r')$ satisfies the inhomogeneous equation,
\begin{equation}
\label{eq:radialr}
\fl
\Big\{\frac{{\rmd}}{{\rmd} r}(r^2-2Mr)\frac{{\rmd}}{{\rmd}r} - \lambda(\lambda+1)-\frac{n^2 \kappa^2 r^4}{r^2-2Mr}\Big\}\chi_{n\lambda}(r,r')=-\frac{1}{\alpha}\delta(r-r').\nonumber\\
\end{equation}

It is convenient to write the radial equation in terms of a new radial variable
$\eta=r/M-1$,
 the radial equation then reads 
\begin{eqnarray}
\label{eq:radial}
\fl
\Big\{\frac{{\rmd}}{{\rmd}\eta}\Big((\eta^{2}-1)\frac{{\rmd}}{{\rmd}\eta}\Big)-\lambda(\lambda+1)-\frac{n^{2}(1+\eta)^{4}}{16(\eta^{2}-1)}\Big\}\chi_{n\lambda}(\eta,\eta')=-\frac{1}{\alpha M}\delta(\eta-\eta'),
\end{eqnarray}
where we have used the fact that $\kappa=1/4M$. For $n=0$, the two solutions of the homogeneous equation are the Legendre functions of the first and second kind. For $n\ne 0$, the homogeneous equation cannot be solved in terms of known functions and must be solved numerically. We denote the two solutions that are regular on the horizon and infinity (or some outer boundary) by $p_{\lambda}^{|n|}(\eta)$ and $q_{\lambda}^{|n|}(\eta)$, respectively. A near-horizon Frobenius analysis for $n \neq 0$ shows that the indicial exponent is $\pm |n|/2$, and so we have the following asymptotic forms:
\begin{equation}
\label{eq:asymp}
\eqalign{
 p_{\lambda}^{|n|}(\eta)\sim (\eta-1)^{|n|/2} \qquad\qquad &\eta\rightarrow 1, \cr
q_{\lambda}^{|n|}(\eta) \sim (\eta-1)^{-|n|/2} &\eta \rightarrow 1.}
\end{equation}
Defining the normalizations by these asymptotic forms and using the Wronskian conditions one can obtain the appropriate normalization of the Green's function:
 \begin{eqnarray}
 \label{eq:chi}
 \chi_{n\lambda}(\eta,\eta') = 
\cases{
\displaystyle{\frac{1}{\alpha M}} P_{\lambda}(\eta_{<})Q_{\lambda}(\eta_{>})&$n=0$, \\
\displaystyle \frac{1}{2|n|\alpha M} p_{\lambda}^{|n|}(\eta_{<}) q_{\lambda}^{|n|}(\eta_{>})\qquad&$n\neq 0$.}
 \end{eqnarray}


\section{Renormalization}
 From the asymptotic forms (\ref{eq:asymp}), it is clear that taking one point on the horizon (with the other outside) means that all the $n\neq 0$ modes vanish. The unrenormalized Green's function then reduces to
\begin{eqnarray}
\label{eq:greensfnnzero}
\fl
G(x,x')=\frac{1}{32 \pi^{2} M^{2} \alpha}&\sum_{m=-\infty}^{\infty} {\rme}^{im(\phi-\phi')} \sum_{l=|m|}^{\infty} (2 \lambda+1) \frac{\Gamma(\lambda+|m|/\alpha+1)}{\Gamma(\lambda-|m|/\alpha+1)} \nonumber\\
& \qquad P^{-|m|/\alpha}_{\lambda}(\cos\theta) P^{-|m|/\alpha}_{\lambda}(\cos\theta') Q_{\lambda}(\eta).
\end{eqnarray}
In the absence of the cosmic string, i.e. $\alpha = 1$, this sum can be done using a combination of the Legendre Addition Theorem and the Heine Identity, yielding a closed form expression~\cite{Candelas:1980zt}, that,
of course, diverges in the coincidence limit. 

To renormalize we follow the spirit of the Christensen-DeWitt\cite{Christensen:1976vb} point-separation method subtracting the geometrical singular
part of the Green's function for small separations in the coordinates. This method relies on the universal Hadamard singularity structure  of the Green's function which ensures that once we have subtracted the geometric singular part, `$U/\sigma + V \ln \sigma$', the coincidence limit of the remainder `$W$' is finite. To calculate $\langle \hat{\varphi}^2 \rangle_{ren}$ in the massless Ricci-flat case, the only subtraction term needed is
\begin{equation}
G_{sing}(x,x') = \frac{1}{4\pi^{2} s^2}
\end{equation}
where $s$ is the geodesic distance between $x$ and $x'$. Having already taken one point ($x'$ say) on the horizon
we choose to separate radially, placing $x$ at $2M+\epsilon$; the geodesic distance is then given by
\begin{eqnarray}
\fl
 s(2M,2M+\epsilon)=\int_{2 M}^{2M+\epsilon} \frac{\rmd r'}{(1-2M/r')^{1/2}}  =  (2M\epsilon)^{1/2} \bigg[2+\frac{\epsilon}{6 M} -\frac{\epsilon^{2}}{80M^{2}}\nonumber\\
 +\frac{\epsilon^{3}}{448 M^{3}}-\frac{5\epsilon^{4}}{9216 M^{4}}+\frac{7\epsilon^{5}}{45056 M^{5}} +\Or(\epsilon^6)\bigg].
 \end{eqnarray}
Note that the leading term behaves like $(\Delta x)^{1/2}$ rather that the usual $\Delta x$ since the metric is singular at 
$r=2M$ so $g^{rr}=O(\epsilon)$, but this is simply a coordinate singularity and in no way affects the validity of the geometrical subtraction.  The subtraction terms up to $\Or(\epsilon)$ are
 \begin{equation}
 \label{eq:gdiv}
 \fl
 G_{sing}  =\frac{1}{32 \pi^{2} M \epsilon} - \frac{1}{192 \pi^{2} M^{2}} + \Or(\epsilon)=\frac{1}{32 \pi^{2} M^{2} (\eta-1)} - \frac{1}{192 \pi^{2} M^{2}} + \Or(\eta-1).
 \end{equation}
 
Taking the partial coincidence limit $\phi \rightarrow \phi'$, $\theta\rightarrow \theta'$ in (\ref{eq:greensfnnzero}), subtracting the renormalization terms (\ref{eq:gdiv}) and taking the limit as $\eta \rightarrow 1$ ($\epsilon\rightarrow 0$) gives us the renormalized expression for the vacuum polarization. However, we now face our fundamental challenge as the limit cannot be taken in a meaningful way in the forms given. In order to take the limit, we must either invent a way to write the divergent terms as an appropriate mode-sum and do a mode-by-mode subtraction or alternatively, we must attempt to perform the sum in (\ref{eq:greensfnhorizon}) so that we can write the Green's function in closed form. In \cite{CSHorizon} we follow the former route to obtain numerical values of $\langle \hat{\varphi}^2 \rangle_{ren}$ outside our black hole. However, when possible, the latter is clearly preferred since it can give us a simple closed-form solution for the renormalized vacuum polarization. In the next section, we derive a formula that will allow us to perform the sum (\ref{eq:greensfnnzero}).


\section{Generalized Heine Identity}
We recall that for real arguments, the Heine Identity for the Legendre Functions is \cite{Erdelyi}
\begin{equation}
\label{eq:heine}
\sum_{l=0}^{\infty} (2 l+1) P_{l}(\Psi)Q_{l}(\zeta)=\frac{1}{(\zeta-\Psi)},
\end{equation}
valid for $|\Psi|<|\zeta|$. Writing $\Psi=\cos\gamma=\cos\theta\cos\theta'+\sin\theta\sin\theta'\cos(\phi-\phi')$, and using the Legendre Addition Theorem,
\begin{equation}
P_{l}(\cos\gamma) = \sum_{m=-l}^{l} \rme^{im (\phi-\phi')} \frac{(l-m)!}{(l+m)!} P_{l}^{m}(\cos\theta)P_{l}^{m}(\cos\theta'),
\end{equation}
we can re-write the Heine Identity as
\begin{equation}
\label{eq:heinenew}
\fl
\sum_{l=0}^{\infty}\sum_{m=-l}^{l} \rme^{im (\phi-\phi')}(2 l+1)\frac{(l-m)!}{(l+m)!} P_{l}^{m}(\cos\theta)P_{l}^{m}(\cos\theta') Q_{l}(\zeta)=\frac{1}{(\zeta-\cos\gamma)}.
\end{equation}
Below we shall generalize this form of the Heine Identity to the non-integer Legendre functions that arise in the mode-sum of a cosmic string space-time. 
It is this form rather than a generalization of (\ref{eq:heine}) that is useful to us since the fundamentaly axially-symmetric nature of the Green's function for $\alpha\neq 1$ means that its radial part  will always depend on $m$ (through $\lambda$), and therefore the $m$-sum over the angular functions can never be done independently of the radial part.

In the Appendix, we have derived other identities involving Legendre Functions of non-integer order and/or degree. We have done so since, although they are not directly useful here, they turn out to be useful in other contexts, for example, determining the Hadamard singularity structure of a 3D Green's function on a cosmic string black hole space-time \cite{CSHorizon}.

Returning to the problem at hand, we want to construct a generalization of (\ref{eq:heinenew}). In addition, we would prefer to do this in such a way that the derivation does not rely on specific properties of the Legendre functions themselves, since
ultimately we wish to apply the same techniques in Kerr space-time. There is a very natural way to do this by equating equivalent expressions for the same Green's function on a conveniently chosen space-time. 
In our case, the most appropriate case to consider is that of  an infinite cosmic string in otherwise flat space-time. Of course, normally in this case one would exploit translational invariance along the 
string and work in cylindrical polar coordinates but for our purposes we choose to work in spherical polar coordinates, so
\begin{equation}
{\rmd}s^{2} = {\rmd}\tau^{2} +{\rmd}r^{2} +r^{2} {\rmd}\theta^{2} +\alpha^{2} r^{2} \sin^{2}\theta {\rmd}\phi^{2},
\end{equation}
where $\phi$ is periodic with period  $2\pi$.
The Green's function for a massless, minimally coupled scalar field on this space-time has a closed-form solution,
which can be obtained, for example, by separating in cylindrical polar coordinates~\cite{Smith}
\begin{equation}
\label{eq:gclosed}
G(x,x') = \frac{1}{8\pi^{2}\alpha} \frac{\sinh(\chi/\alpha)}{\rho\rho' \sinh\chi(\cosh(\chi/\alpha)-\cos(\phi-\phi'))},
\end{equation}
where
\begin{equation}
\cosh\chi = \frac{(\tau-\tau')^{2}+r^2 +r'^2-2r r' \cos\theta\cos\theta'}{2\rho\rho'}
\end{equation}
and $\rho=r\sin\theta$ is the cylindrical polar coordinate. 

One can also obtain the spherical polar mode-sum expression for this Green's function:
\begin{eqnarray}
\fl
G(x,x') = \frac{1}{4\pi^{2}\alpha} \int_{0}^{\infty}\cos\omega(\tau-\tau') \rmd\omega \sum_{m=-\infty}^{\infty} \rme^{im(\phi-\phi')} \sum_{l=|m|}^{\infty} (2 \lambda+1) \frac{\Gamma(\lambda+|m|/\alpha+1)}{\Gamma(\lambda-|m|/\alpha+1)}  \nonumber\\
 P^{-|m|/\alpha}_{\lambda}(\cos\theta) P^{-|m|/\alpha}_{\lambda}(\cos\theta') R_{\omega\lambda}(r,r'),
\end{eqnarray}
where $R_{\omega\lambda}(r,r')$ is the 1D Green's function satisfying
\begin{equation}
\Big[\frac{{\rmd}}{{\rmd} r}\Big(r^{2} \frac{{\rmd}}{{\rmd} r}\Big)-\omega^{2} r^{2} -\lambda(\lambda+1)\Big]R_{\omega\lambda} = -\delta(r-r').
\end{equation}
The solution of this equation can be given in terms of Modified Bessel Functions, 
\begin{equation}
R_{\omega\lambda}(r,r') = \frac{I_{\lambda+1/2}(\omega r_{<}) K_{\lambda +1/2}(\omega r_{>})}{(r r')^{1/2}},
\end{equation}
where $I$ and $K$ are the Modified Bessel Functions of the first and second kinds, respectively.
The integral over $\omega$ can now be performed using the identity \cite{gradriz}
\begin{equation}
\int_{0}^{\infty} \cos\omega(\tau-\tau') I_{\lambda+1/2}(\omega r_{<}) K_{\lambda +1/2}(\omega r_{>})\rmd\omega = \frac{1}{2(r r')^{1/2}} Q_{\lambda}(\zeta)
\end{equation}
valid for $\lambda > -1$, where $Q_\lambda$ is the Legendre function of the second kind and
\begin{equation}
\zeta = \frac{(\tau-\tau')^{2} + r^{2} +r'^{2}}{2 r r'}.
\end{equation}
Thus, the spherical polar mode-sum expression for the Green's Function may be written as
\begin{eqnarray}
\fl
G(x,x') = \frac{1}{8 \pi^{2}\alpha r r'} \sum_{m=-\infty}^{\infty} e^{im(\phi-\phi')} \sum_{l=|m|}^{\infty} (2 \lambda+1) \frac{\Gamma(\lambda+|m|/\alpha+1)}{\Gamma(\lambda-|m|/\alpha+1)} \nonumber\\
 P^{-|m|/\alpha}_{\lambda}(\cos\theta) P^{-|m|/\alpha}_{\lambda}(\cos\theta') Q_{\lambda}(\zeta).
\end{eqnarray}

We now equate this Green's Function with the equivalent closed form expression (\ref{eq:gclosed}) to obtain the following generalized Heine Identity:
\begin{eqnarray}
\label{eq:generalizedheine}
\fl
\sum_{m=-\infty}^{\infty} \rme^{im(\phi-\phi')} \sum_{l=|m|}^{\infty} (2 \lambda+1) \frac{\Gamma(\lambda+|m|/\alpha+1)}{\Gamma(\lambda-|m|/\alpha+1)}P^{-|m|/\alpha}_{\lambda}(\cos\theta) P^{-|m|/\alpha}_{\lambda}(\cos\theta')Q_{\lambda}(\zeta)  \nonumber\\
=\frac{\sinh(\chi/\alpha)}{\sin\theta\sin\theta' \sinh\chi(\cosh(\chi/\alpha)-\cos(\phi-\phi'))}
\end{eqnarray}
where
\begin{equation}
\fl
\cosh\chi = \frac{\zeta-\cos\theta\cos\theta'}{\sin\theta\sin\theta'}\quad \Leftrightarrow \quad
\zeta = \cos \gamma + (\cosh\chi - \cos (\phi-\phi')) \sin\theta\sin\theta'.
\end{equation}
This identity, valid for $\zeta>|\cos \gamma|$, is completely analagous to (\ref{eq:heinenew}) and reduces to it in the $\alpha\rightarrow 1$ limit.


\section{Calculation of $\langle \hat{\varphi}^{2} \rangle_{ren}$}
Returning now to our calculation of the vacuum polarization on the horizon of the Schwarzschild black hole threaded by a cosmic string, we see that our generalized Heine Identity (\ref{eq:generalizedheine}) has precisely the right form to enable us to perform the mode-sum in the Green's function expression (\ref{eq:greensfn}). The Green's function, where one point has been taken to lie on the horizon, may now be written as
\begin{equation}
\label{eq:greensfnhorizon}
G(x,x')=\frac{1}{32\pi^{2} M^{2}\alpha}\frac{\sinh(\chi/\alpha)}{\sin\theta\sin\theta' \sinh\chi(\cosh(\chi/\alpha)-\cos(\phi-\phi'))}
\end{equation}
For purely radial separation, we take $\phi'\rightarrow\phi$, $\theta'\rightarrow\theta$, to yield
\begin{equation}
\label{eq:gclosedradial}
G(r,\theta;2M,\theta)=\frac{1}{32\pi^{2} M^{2}\alpha}\frac{\sinh(\chi/\alpha)}{\sin^{2}\theta\sinh\chi(\cosh(\chi/\alpha)-1)}
\end{equation}
where we now have
\begin{equation}
\cosh\chi=1+\frac{(\eta-1)}{\sin^{2}\theta}.
\end{equation}
From Eqs.(\ref{eq:gdiv}) and (\ref{eq:gclosedradial}), the renormalized vacuum polarization is given by
\begin{eqnarray}
\label{eq:phi2limit}
&&\fl\langle \hat{\varphi}^{2} \rangle_{ren}^{horizon} = \lim_{\eta \rightarrow 1} \Big[ G(r,\theta;2M,\theta)-G_{sing}(r;2M)\Big] \nonumber\\
&&\fl \ = \lim_{\eta \rightarrow 1} \Big[ \frac{1}{32\pi^{2} M^{2} \alpha \sin^{2}\theta} \frac{\sinh(\chi/\alpha)}{\sinh\chi(\cosh(\chi/\alpha)-1)}-\frac{1}{32\pi^{2} M^{2} (\eta-1)}-\frac{1}{192\pi^{2} M^{2}}\Big].
\end{eqnarray}
It is now convenient first expand in terms of $\chi$
 \begin{equation}
  \frac{\sinh(\chi/\alpha)}{\sinh\chi(\cosh(\chi/\alpha)-1)}=\frac{2\alpha}{\chi^{2}} +\Big(\frac{1}{6\alpha}-\frac{\alpha}{3}\Big) + \Or(\chi^{2}).
  \end{equation}
then 
  \begin{equation}
  \cosh\chi = 1+\frac{(\eta-1)}{\sin^{2}\theta}\quad \Rightarrow \quad \frac{1}{\chi^{2}} = \frac{\sin^{2}\theta}{2 (\eta-1)} + \frac{1}{12} +\Or(\eta-1)^{2}
  \end{equation}
so we  obtain
\begin{equation}
\label{eq:gexpansion}
 \frac{\sinh(\chi/\alpha)}{\sinh\chi(\cosh(\chi/\alpha)-1)}=\frac{\alpha \sin^{2}\theta}{(\eta-1)} +\frac{1}{6\alpha}(1-\alpha^{2}) + \Or(\eta-1)^{2}.
 \end{equation}
Substituting (\ref{eq:gexpansion}) into (\ref{eq:phi2limit}) and taking the limit, we arrive at the renormalized vacuum polarization on the horizon,
\begin{equation}
\label{eq:phi2ren}
\langle \hat{\varphi}^{2} \rangle_{ren}^{horizon} = \frac{1}{192\pi^{2} M^{2}}\Big(1+\frac{1-\alpha^{2}}{\alpha^{2}\sin^{2}\theta}\Big).
  \end{equation}
  
  The first term here is simply the Candelas result \cite{Candelas:1980zt}, which, of course we recover when $\alpha \rightarrow 1$, and the second term represents the contribution due to the presence of the cosmic string. The result (\ref{eq:phi2ren}) has in fact been calculated by Davies and Sahni \cite{DaviesSahni} but for a restricted set of $\alpha$ values such that $(1/\alpha)$ is an integer. For this restricted class of values, the cosmic string solution is obtained as a sum of $(1/\alpha)$ images of the solution in the absence of a string. The method of images cannot be applied for general azimuthal deficits, however, so our method presents a generalization and independent derivation of the result of Davies and Sahni.
  
\begin{figure}
\centering
\includegraphics[width=11cm]{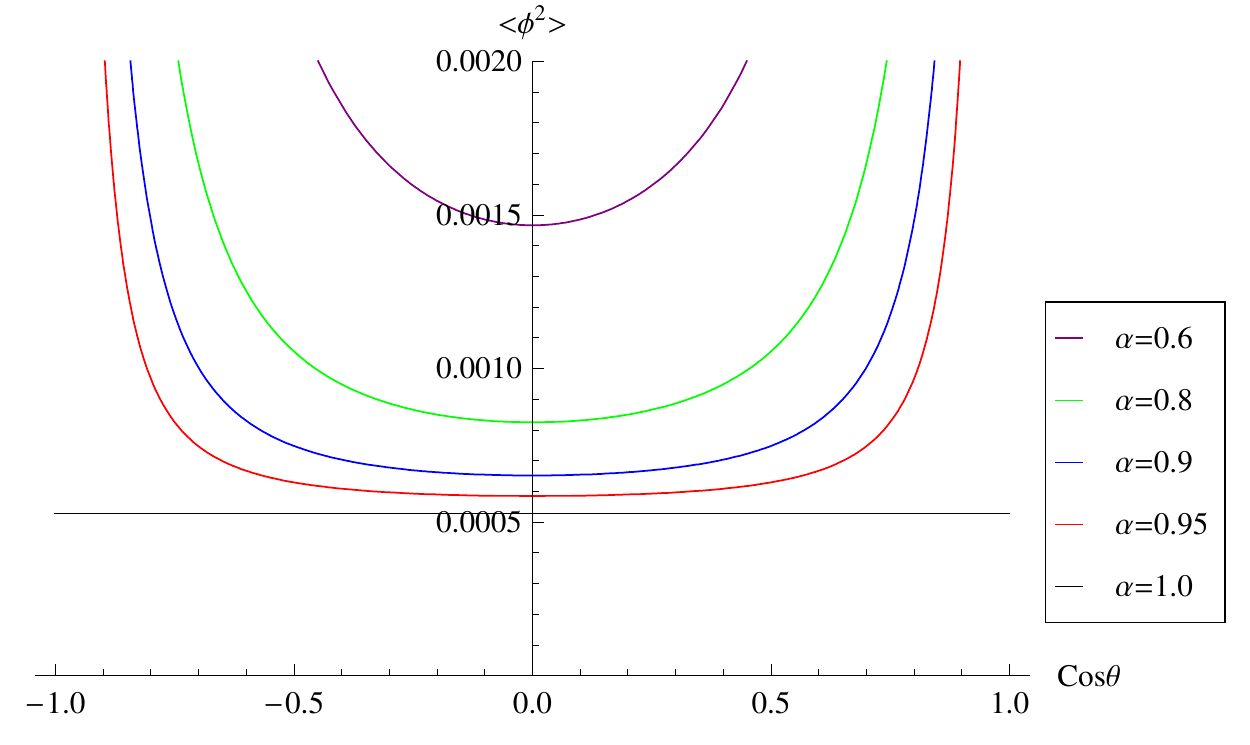}
\caption{\emph{ $\langle \hat{\varphi}^{2} \rangle_{ren}^{horizon}$} (in units $M=1$) as a function of $\cos\theta$ for a range of values of $\alpha$. For $\alpha \neq 1$,  $\langle \hat{\varphi}^{2} \rangle_{ren}^{horizon}$  diverges at the poles since there is a curvature singularity there due to the presence of the string. $\langle \hat{\varphi}^{2} \rangle_{ren}^{horizon}$ increases everywhere as $\alpha$ decreases.}
\label{fig:phi2horizon}
\end{figure}

  We have plotted the vacuum polarization in Figure~\ref{fig:phi2horizon} for a range of values of $\alpha$. We see that the presence of the string increases $\langle \hat{\varphi}^{2} \rangle_{ren}^{horizon}$ everywhere. In the equatorial plane it is increased by a factor of $1/\alpha^2$ from its value in the absence of a string.
As we approach the axis  $\langle \hat{\varphi}^{2} \rangle_{ren}^{horizon}$ diverges in a non-integrable manner as 
 \begin{equation}
\langle \hat{\varphi}^{2} \rangle_{ren}^{horizon} \sim \frac{1}{48\pi^{2} }\frac{(1-\alpha^{2})}{\alpha^{2}} \frac{1}{(2M \sin\theta)^{2}} \qquad \sin\theta \to 0,
\end{equation}
  as one would expect a distance  $(2M \sin\theta)$ from a flat-space cosmic string.
  The region over which the string dominates may be characterised by the range of $\cos \theta$ (near $\theta=0$) at which 
  $\langle \hat{\varphi}^{2} \rangle_{ren}^{horizon}$ exceeds twice its value in the equatorial plane, this is given
by
  \begin{equation}
1-(\cos \theta)_2 = 1-\frac{1}{(2-\alpha^2)^{1/2}} = (1-\alpha) + O\bigl((1-\alpha)^2)\bigr) .
\end{equation}

\section{The Stress Tensor for a Massless Scalar Field with Arbitrary Coupling}
In this section, we shall employ our generalized Heine identity (\ref{eq:heine}) in order to calculate the renormalized expectation value of the stress energy tensor for a massless scalar field with arbitrarily coupling to the Ricci scalar in the Hartle-Hawking vacuum state. This quantity is of particular physical interest in obtaining the back-reaction on the spacetime geometry in the semi-classical approximation to quantum gravity.

We first draw some general conclusions about the quantity we wish to calculate. We know that, by definition, the Hartle-Hawking vacuum respects the symmetries of the spacetime and that the metric given by Eq.(\ref{eq:metric}) is symmetric under the transformations
\begin{equation}
\tau \rightarrow -\tau  \qquad\qquad \phi\rightarrow-\phi  \qquad\qquad \theta\rightarrow \pi-\theta ,
\end{equation}
and under the  translational symmetries
\begin{equation}
\tau \rightarrow \tau+\tau_{0}  \qquad\qquad \phi\rightarrow\phi+\phi_{0}.
\end{equation}
We note, however, that unlike the spherically symmetric case, we do not have a translational symmetry in the $\theta$-direction, but only the discrete reflective symmetry in the equatorial plane. As such, we have the possibility of non-zero off-diagonal components $T^{r}{}_{\theta}$ and $T^{\theta}{}_{r}$. All other off-diagonal components are identically zero for the Hartle-Hawking vacuum. For a massless scalar field propagating in a Ricci-flat background, we have the following expression for the classical stress tensor
\begin{eqnarray}
\label{eq:Tclassical}
\fl
\qquad T^{a}{}_{b} =(1-2\xi) g^{a c}\varphi_{;c}\varphi_{;b}+(2\xi-\case{1}{2})\delta^{a}{}_{b}g^{c d}\varphi_{;c}\varphi_{;d}-2\xi g^{a c}\varphi \varphi_{;c b}+2\xi\delta^{a}{}_{b}\varphi\Box\varphi
\end{eqnarray}
It will prove convenient to write this tensor as
\begin{equation}
T^{a}{}_{b} = [\hat{\tau}^{a}{}_{b}(\varphi(x)\varphi(x'))]\equiv \lim_{x'\rightarrow x} \hat{\tau}^{a}{}_{b}(\varphi(x)\varphi(x'))
\end{equation}
where $\hat{\tau}^{a}{}_{b}=\hat{\tau}^{a}{}_{ b}(x,x')$ is a differential operator which may be defined in any way provided it gives (\ref{eq:Tclassical}) in the coincidence limit. We shall adopt the following definition,
\begin{equation}
\label{eq:stresstensorop}
\fl
\qquad \hat{\tau}^{a}{}_{ b} = (1-2\xi) g^{a c'}\nabla_{b}\nabla_{c'}+(2\xi-\case{1}{2})\delta^{a}{}_{b}g^{c d'}\nabla_{c}\nabla_{d'}-2\xi g^{a c}\nabla_{c}\nabla_{b}+2\xi\delta^{a}{}_{b}\nabla_{c}\nabla^{c}
\end{equation}
where $g^{a b'}$ are the bivectors of parallel transport. Using point separation,  the regularized, unrenormalized quantum stress tensor in the Hartle-Hawking vacuum state arises when this differential operator acts on the thermal Euclidean Green's function (\ref{eq:greensfn}),
\begin{eqnarray}
\label{eq:Tunren}
\fl
\qquad \langle \hat{T}^{a}{}_{b} \rangle_{unren} = (1-2\xi) g^{a c'}G_{;c' b}+(2\xi-\case{1}{2})\delta^{a}{}_{b}g^{c d'}G_{;c d'} -2\xi g^{a c}G_{;c b}+2\xi\delta^{a}{}_{b}\Box G.
\end{eqnarray}
This is clearly divergent in the coincident limit and must be renormalized prior to relaxing the regularization. To achieve this we note that this Green's function is of Hadamard form and may be written as
\begin{equation}
\label{eq:hadamard}
G(x,x') = \frac{1}{8\pi^{2}}\left(\frac{\Delta^{1/2}}{\sigma}+V\ln(\sigma)+W\right)
\end{equation}
where $\sigma(x,x')$ is half the square of the geodesic distance between $x$ and $x'$ and
\begin{equation}
\Delta=-g^{-1/2}(x)\det(\sigma_{;a b'})g^{-1/2}(x')
\end{equation}
is the Van Vleck-Morette determinant. $V(x,x')$ and $W(x,x')$ are regular symmetric biscalar functions with $V(x,x)$ being
geometrical (local) while the determination of $W(x,x')$ requires (global) boundary conditions.  Thus  $V(x,x)$ captures the 
(local) ultraviolet divergences of the quantum  theory while $W(x,x')$  captures the state dependence.

The divergent part of the stress tensor is then the result of the operator (\ref{eq:stresstensorop}) acting on the singular part of the Green's function, i.e.
\begin{equation}
\label{eq:tdivhadamard}
\langle \hat{T}^{a}{}_{ b} \rangle_{div} = \frac{1}{8\pi^{2}} \hat{\tau}^{a}{}_{ b} \left( \frac{\Delta^{1/2}}{\sigma}+V\ln(\sigma)\right).
\end{equation}
The renormalized stress tensor then essentially has to be given by the coincident limit of the differential operator (\ref{eq:stresstensorop}) acting on the regular part $W$,
\begin{equation}
\label{eq:tW}
 \frac{1}{8\pi^{2}} \left[\hat{\tau}^{a}{}_{ b} W(x,x')\right].
 \end{equation}
Rather than calculating $W(x,x')$ itself, we calculate (\ref{eq:tW}) by subtracting Eq.(\ref{eq:tdivhadamard}) from (\ref{eq:Tunren}) and then 
take thing coincidence. There is one caveat, however, since the application of the differential operator $\hat{\tau}^{a}{}_{ b}$ to $W(x,x')$ results in a stress tensor which is not conserved since $W(x,x')$ is not a solution to the homogeneous wave equation but rather an inhomogeneous one where the source is geometrical. It is easy to redeem this situation by adding a regular geometrical term to the definition of the stress tensor \cite{BrownOttewill1986},
\begin{equation}
\label{eq:tren}
\langle \hat{T}^{a}{}_{ b} \rangle_{ren}= \left[\langle \hat{T}^{a}{}_{ b} \rangle_{unren}-\langle \hat{T}^{a}{}_{ b} \rangle_{div}\right]+\frac{1}{4\pi^{2}}v_{1}\delta^{a}{}_{ b} 
\end{equation}
where for a massless scalar field in a Ricci-flat background, the geometrical scalar $v_{1}$ has the very simple form
\begin{equation}
\label{eq:v1}
v_{1}=\frac{1}{720}C_{p q r s}C^{p q r s} .
\end{equation}
On the horizon of our spacetime this reduces to $v_1(2M)={\kappa^{2}}/({60 M^{2}}) $.

In this paper, we are concerned with the calculation of the stress tensor on the horizon only. In this case, it is most convenient to split in the radial direction, taking the inner point (say $x'$), to lie on the horizon. Where derivatives of the Green's function are concerned, we must first perform the derivatives before we take partial coincidence limits. The bivectors of parallel transport have been calculated for radial separation in reference \cite{Candelas:1980zt} for the Schwarzschild spacetime and are trivially extended to the cosmic string case by multiplying the azimuthal component by an $\alpha^{2}$ factor. We list them here for completeness:
\begin{eqnarray*}
\fl
g_{\tau\tau'}=(1-2M/r)^{1/2}(1-2M/r')^{1/2} &\qquad
g_{r r'}=(1-2M/r)^{-1/2}(1-2M/r')^{-1/2} \nonumber\\
\fl g_{\theta\theta'}= r r'  &\qquad
g_{\phi\phi'} = r r' \alpha^{2} \sin^{2}\theta.
\end{eqnarray*}


\subsection{The Derivatives of the Green's Function}
Examination of (\ref{eq:Tunren}) reveals that there is essentially two types of terms we need to evaluate: those of the form $g^{a c'}G_{;c' b}$ and those of the form $g^{a c}G_{;c b}$. 

Considering the former case first, such terms involve a covariant derivative at each spacetime point but since $G(x,x')$ is a scalar at both $x$ and $x'$, we are in fact only dealing with partial derivatives. For the angular terms $g^{\theta\theta'}G_{;\theta\theta'}$ and $g^{\phi\phi'}G_{;\phi\phi'}$, it is clear from the asymptotic forms (\ref{eq:asymp}) that taking $x'$ to lie on the horizon means that only the $n=0$ terms will contribute in the limit $\eta'\rightarrow 1$. Therefore, we can differentiate directly Eq.(\ref{eq:greensfnhorizon}) which is straight forward. Taking partial coincidence limits and expanding about the horizon, we obtain
\begin{eqnarray}
\label{eq:Gthetathetap}
\fl
[g^{\theta\theta'} G_{;\theta\theta'}]_{\mathrm{r}}=\frac{\kappa^{2}}{8\pi^{2}M^{2}(\eta-1)^{2}}-\frac{\kappa^{2}}{16 \pi^{2} M^{2} (\eta-1)}+\frac{\kappa^{2}}{32 \pi^{2} M^{2}} \nonumber\\
-\frac{\kappa^{2}(\alpha^{2}-1)}{1440\pi^{2} M^{2}} \Big(\frac{(11\alpha^{2}+1)+30\alpha^{2}\cos^{2}\theta}{\alpha^{4}\sin^{4}\theta}\Big)+\Or(\eta-1)
\end{eqnarray}
and
\begin{eqnarray}
\label{eq:Gphiphip}
\fl
[g^{\phi\phi'} G_{;\phi\phi'}]_{\mathrm{r}}=\frac{\kappa^{2}}{8\pi^{2}M^{2}(\eta-1)^{2}}-\frac{\kappa^{2}}{16 \pi^{2} M^{2} (\eta-1)}+\frac{\kappa^{2}}{32 \pi^{2} M^{2}} \nonumber\\
\qquad+\frac{\kappa^{2}}{480 \pi^{2} M^{2}} \frac{(\alpha^{2}-1)(\alpha^{2}+1)}{\alpha^{4}\sin^{4}\theta} +\Or(\eta-1),
\end{eqnarray}
where we have adopted square bracket notation $[..]_{\mathrm{r}}$ to indicate that we have taken the partial coincidence limit $(\tau\rightarrow \tau',\eta'\rightarrow 1,\theta \rightarrow \theta',\phi\rightarrow \phi')$.

For $g^{\tau\tau'}G_{;\tau\tau'}$ and $g^{r r'}G_{;r r'}$, we must differentiate the full Green's function given by Eqs.(\ref{eq:greensfn}) and (\ref{eq:chi}) before we can take $x'$ to lie on the horizon. Considering $g^{\tau\tau'}G_{;\tau\tau'}$ first, we have
\begin{eqnarray}
\fl
g^{\tau\tau'}G_{;\tau\tau'} = \frac{1}{(1-2M/r)^{1/2}(1-2M/r')^{1/2}}\frac{\kappa^{3}}{8\pi^{2}}\sum_{n=-\infty}^{\infty} n^{2}{\rme}^{i n \kappa (\tau-\tau')}\sum_{m=-\infty}^{\infty} {\rme}^{im(\phi-\phi')} \nonumber\\
\sum_{l=|m|}^{\infty} (2\lambda+1)\frac{\Gamma(\lambda+|m|/\alpha+1)}{\Gamma(\lambda-|m|/\alpha+1)} P_{\lambda}^{-|m|/\alpha}(\cos\theta) P_{\lambda}^{-|m|/\alpha}(\cos\theta')\chi_{n\lambda}(\eta,\eta')
 \end{eqnarray}
 Trivially, the $n=0$ term will vanish. Moreover, using the asymptotic forms Eq.(\ref{eq:asymp}), it is clear that taking $x'$ to the horizon only the terms corresponding $n=\pm 1$ survive. We now have, in the partial coincidence limit,
 \begin{eqnarray}
 \label{eq:Gtautaup}
 \fl
 [g^{\tau\tau'} G_{;\tau\tau'}]_{\mathrm{r}} = \frac{\kappa^{2}}{16 \pi^{2}M^{2}\alpha} \frac{(\eta+1)^{1/2}}{2^{1/2}
	(\eta-1)^{1/2}} \sum_{m=-\infty}^{\infty}\sum_{l=|m|}^{\infty} (2\lambda+1)\frac{\Gamma(\lambda+|m|/\alpha+1)}{\Gamma(\lambda-|m|/\alpha+1)} \nonumber\\
 \end{eqnarray}
 We will denote the double sum by
 \begin{equation}
 \label{eq:fdefn}
 \fl
 F(\eta,\cos\theta)\equiv\sum_{m=-\infty}^{\infty}\sum_{l=|m|}^{\infty} (2\lambda+1)\frac{\Gamma(\lambda+|m|/\alpha+1)}{\Gamma(\lambda-|m|/\alpha+1)}[ P_{\lambda}^{-|m|/\alpha}(\cos\theta)]^{2} q_{1\lambda}(\eta) ,
 \end{equation}
 and return to its evaluation later; this will involve both numerical and analytic parts.
 
 We follow a similar argument to obtain an expression for $g^{rr'}G_{;rr'}$. We have
 \begin{eqnarray}
 \fl
 g^{rr'}G_{;rr'} = (1-2M/r)^{1/2}(1-2M/r')^{1/2}\frac{\kappa}{8\pi^{2}}\sum_{n=-\infty}^{\infty}{\rme}^{i n \kappa (\tau-\tau')}\sum_{m=-\infty}^{\infty} {\rme}^{im(\phi-\phi')} \nonumber\\
\hspace{-2cm} \times\sum_{l=|m|}^{\infty} (2\lambda+1)\frac{\Gamma(\lambda+|m|/\alpha+1)}{\Gamma(\lambda-|m|/\alpha+1)} P_{\lambda}^{-|m|/\alpha}(\cos\theta) P_{\lambda}^{-|m|/\alpha}(\cos\theta')\frac{\partial^{2}}{\partial \eta \partial \eta'}\chi_{n\lambda}(\eta,\eta'). 
 \end{eqnarray}
 Splitting up the $n=0$ and $n\ne 0$ terms and writing in terms of radial variable $\eta$, we have
 \begin{eqnarray}
 \fl
  g^{rr'}G_{;rr'} =\frac{ (\eta-1)^{1/2}(\eta'-1)^{1/2}}{(\eta+1)^{1/2}(\eta'+1)^{1/2}}\frac{\kappa}{8\pi^{2}\alpha M^{3}}\sum_{m=-\infty}^{\infty} {\rme}^{im(\phi-\phi')} \sum_{l=|m|}^{\infty} (2\lambda+1)\frac{\Gamma(\lambda+|m|/\alpha+1)}{\Gamma(\lambda-|m|/\alpha+1)} \nonumber\\
\hspace{-2cm} \times P_{\lambda}^{-|m|/\alpha}(\cos\theta) P_{\lambda}^{-|m|/\alpha}(\cos\theta')\left(\frac{\rmd P_{\lambda}(\eta')}{\rmd \eta'}\frac{\rmd Q_{\lambda}(\eta)}{\rmd \eta} +\sum_{n=1}^{\infty}{\rme}^{i n \kappa(\tau-\tau')}\frac{\rmd p_{n\lambda}(\eta')}{\rmd \eta'}\frac{\rmd q_{n\lambda}(\eta)}{\rmd \eta}\right).\nonumber\\
\end{eqnarray}
Now using the fact that
\begin{equation}
(\eta'^{2}-1)^{1/2}\frac{\rmd P_{\lambda}(\eta')}{\rmd \eta'}=P^{1}_{\lambda}(\eta') \rightarrow 0 \qquad \mathrm{as} \qquad \eta'\rightarrow1
\end{equation}
implies that the $n=0$ term vanishes. Also, using the asymptotic forms Eq.(\ref{eq:asymp}), we have for $n\ne 0$,
\begin{equation}
(\eta'-1)^{1/2}\frac{1}{|n|}\frac{\rmd p_{n\lambda}(\eta')}{\rmd \eta'}=\frac{1}{2}(\eta'-1)^{|n|/2-1/2}
\end{equation}
which will vanish for all but the $n=\pm1$ terms in the $\eta'\rightarrow 1$ limit. Therefore, the only contribution to $g^{rr'}G_{rr'}$ comes from the $n=\pm1$ terms, and in the partial coincidence limit we have
\begin{eqnarray}
\label{eq:Grrp}
\fl
[g^{rr'}G_{;rr'}]_{\mathrm{r}}=\frac{\kappa^{2}}{4 \pi^{2}M^{2}\alpha}\frac{(\eta-1)^{1/2}}{2^{1/2}(\eta+1)^{1/2}}\nonumber\\ \times\sum_{m=-\infty}^{\infty}\sum_{l=|m|}^{\infty} (2\lambda+1)\frac{\Gamma(\lambda+|m|/\alpha+1)}{\Gamma(\lambda-|m|/\alpha+1)} 
 [ P_{\lambda}^{-|m|/\alpha}(\cos\theta)]^{2} \frac{\rmd q_{1\lambda}(\eta)}{\rmd \eta}  \nonumber\\
 \fl \qquad\qquad \,\,\,=\frac{\kappa^{2}}{4\pi^{2}M^{2}\alpha} \frac{(\eta-1)^{1/2}}{2^{1/2}(\eta+1)^{1/2}}\frac{\rmd F(\eta,\cos\theta)}{\rmd \eta}.
 \end{eqnarray}
 
 Finally, we require $g^{rr'}G_{;r'\theta}$. Similar calculations to those above show that only the $n=\pm 1$ term survives when we differentiate with respect to $r'$ and then take this point to lie on the horizon, yielding
 \begin{eqnarray}
 \fl
 g^{rr'}G_{;r'\theta}=\frac{\kappa^{2}}{4\pi^{2} M \alpha}\frac{(\eta-1)^{1/2}}{2^{1/2}(\eta+1)^{1/2}}
 \nonumber\\
\hspace{-2cm}\times\frac{\partial}{\partial\theta}\sum_{m=-\infty}^{\infty}\sum_{l=|m|}^{\infty} (2\lambda+1)\frac{\Gamma(\lambda+|m|/\alpha+1)}{\Gamma(\lambda-|m|/\alpha+1)}
 P_{\lambda}^{-|m|/\alpha}(\cos\theta) P_{\lambda}^{-|m|/\alpha}(\cos\theta') q_{1\lambda}(\eta).
 \end{eqnarray}
 In order to take the $\theta\rightarrow\theta'$ limit, we make use of the fact that
 \begin{eqnarray}
\label{eq:thetalimit}
\fl
\lim_{\theta\rightarrow\theta'}\Big[\sum_{m=-\infty}^{\infty}\sum_{l=|m|}^{\infty}(2\lambda+1)\frac{\Gamma(\lambda+|m|/\alpha+1)}{\Gamma(\lambda-|m|/\alpha+1)}\Big(\frac{\partial}{\partial \theta}P^{-|m|/\alpha}_{\lambda}(\cos\theta)\Big)P^{-|m|/\alpha}_{\lambda}(\cos\theta')\Big]\nonumber\\
=\frac{1}{2}\frac{\partial}{\partial\theta}\Big(\sum_{l=|m|}^{\infty}(2\lambda+1)\frac{\Gamma(\lambda+|m|/\alpha+1)}{\Gamma(\lambda-|m|/\alpha+1)}P^{-|m|/\alpha}_{\lambda}(\cos\theta)^{2}\Big).
\end{eqnarray}
Therefore, in the partial coincidence limit, we have
\begin{eqnarray}
\label{eq:Grptheta}
[g^{rr'}G_{;r'\theta}]_{\mathrm{r}}=\frac{\kappa^{2}}{8\pi^{2} M \alpha}\frac{(\eta-1)^{1/2}}{2^{1/2} (\eta+1)^{1/2}}\frac{\partial}{\partial\theta}F(\eta,\cos\theta).
\end{eqnarray} 

We now consider derivatives of the form $g^{a c}G_{;c b}$. For such terms, we have two derivatives at the same spacetime point, which will involve the Christoffel symbols,
\begin{equation}
\label{eq:gcovariant}
G_{;a b} = G_{, a b}-\Gamma^{c}_{a b} G_{,c},
\end{equation}
where the Christoffel symbols are the same as those of Schwarzschild spacetime except
\begin{eqnarray}
\label{eq:christoffel}
 \Gamma^{r}_{\phi\phi}=-\alpha^{2} (r-2M)\sin^{2}\theta &\qquad
\Gamma^{\theta}_{\phi\phi}=-\alpha^{2}\sin\theta\cos\theta \ .
\end{eqnarray}
 In spite of this minor added complication, things are significantly easier in this case since only the $n=0$ term will contribute for each derivative. This is because we can always choose the differentiation to act on the outer point, this will not affect the asymptotic structure at the other spacetime point and so taking $\eta'\rightarrow 1$ and using the asymptotic forms (\ref{eq:asymp}), it is clear that only the $n=0$ term will contribute in the limit. Now, the derivatives in (\ref{eq:gcovariant}) can be obtained by direct differentiation of (\ref{eq:greensfnhorizon}). Performing the derivatives and using the appropriate Christoffel symbols, followed by taking the partial coincidence limits and expanding about the horizon, we obtain
  \begin{eqnarray}
  \label{eq:Gtautau}
 \fl
\ \ [g^{\tau\tau}G_{;\tau\tau}]_{\mathrm{r}}=-\frac{\kappa^{2}}{8\pi^{2}M^{2}(\eta-1)^{2}}+\frac{\kappa^{2}}{8\pi^{2}M^{2}(\eta-1)}-\frac{3\kappa^{2}}{32\pi^{2}M^{2}}\nonumber\\
 \qquad+\frac{\kappa^{2}}{1440\pi^{2}M^{2}}\frac{(\alpha^2-1)(11 \alpha^2 +1)}{\alpha^{4}\sin^{4}\theta} +\Or(\eta-1),
 \end{eqnarray}
 \begin{eqnarray}
 \label{eq:Grr}
 \fl
\ \ [g^{r r}G_{;rr}]_{\mathrm{r}}=\frac{3\kappa^{2}}{8\pi^{2}M^{2}(\eta-1)^{2}}-\frac{\kappa^{2}}{8\pi^{2}M^{2}(\eta-1)}+\frac{\kappa^{2}}{32\pi^{2}M^{2}}\nonumber\\
 \qquad+\frac{\kappa^{2}}{1440\pi^{2}M^{2}}\frac{(\alpha^2-1)(11 \alpha^2 +1) }{\alpha^{4}\sin^{4}\theta} +\Or(\eta-1),
 \end{eqnarray}
  \begin{eqnarray}
  \label{eq:Gthetatheta}
 \fl
\ \ [g^{\theta\theta}G_{;\theta\theta}]_{\mathrm{r}}=-\frac{\kappa^{2}}{8\pi^{2}M^{2}(\eta-1)^{2}}+\frac{\kappa^{2}}{32\pi^{2}M^{2}}+\frac{\kappa^{2}}{48\pi^{2}M^{2}}\frac{(\alpha^{2}-1)}{\alpha^{2}\sin^{2}\theta}\nonumber\\
 \qquad-\frac{\kappa^{2}}{1440\pi^{2}M^{2}}\frac{(\alpha^2-1)(49 \alpha^2 -1)}{\alpha^{4}\sin^{4}\theta} +\Or(\eta-1),
 \end{eqnarray}
 \begin{eqnarray}
   \label{eq:Gphiphi}
 \fl
\ \ [g^{\phi\phi}G_{;\phi\phi}]_{\mathrm{r}}=-\frac{\kappa^{2}}{8\pi^{2}M^{2}(\eta-1)^{2}}+\frac{\kappa^{2}}{32\pi^{2}M^{2}}-\frac{\kappa^{2}}{480\pi^{2}M^{2}}\frac{(\alpha^{2}-1)(\alpha^{2}+1)}{\alpha^{4}\sin^{4}\theta} \nonumber\\
\qquad+\frac{\kappa^{2}}{48\pi^{2}M^{2}}\frac{(\alpha^{2}-1)}{\alpha^{2}}\frac{\cos^{2}\theta}{\sin^{4}\theta}+\Or(\eta-1),
 \end{eqnarray}
 \begin{eqnarray}
 \label{eq:Grtheta}
 \fl
\ \ [g^{rr}G_{;r\theta}]_{\mathrm{r}}=-\frac{\kappa^{2}}{720\pi^{2}M^{3}}
\frac{(\alpha^2-1)(11 \alpha^2 +1)\cos\theta}{\alpha^{4}\sin^{5}\theta}-
\frac{\kappa^{2}}{96\pi^{2}M^{3}}\frac{(\alpha^{2}-1)\cos\theta}{\alpha^{2}\sin^{3}\theta}+\Or(\eta-1). \nonumber\\
\end{eqnarray}
 
 
 \subsection{Candelas Method for Obtaining a Series Solution for $F(\eta,\cos\theta)$}
 In order to proceed with the evaluation of the stress tensor on the horizon, we require a series solution for the function we have called $F(\eta,\cos\theta)$ (\ref{eq:fdefn}) about the horizon $\eta=1$. Fortunately, Candelas \cite{Candelas:1980zt} has obtained the series solution in the Schwarzschild case and the method described here follows an almost identical path, so we refer the reader to the Candelas paper for more details.
Following Candelas, we may capture the asymptotic behaviour of $q_{1\lambda}(\eta)$ as $\eta  \to 1$ by writing
 \begin{equation}
 \label{eq:q1def}
 q_{1\lambda}(\eta) = -\sqrt{2}\,Q^{1}_{\lambda}(\eta)+\beta_{\lambda} p_{1\lambda}(\eta)-f(\eta) Q_{\lambda}(\eta)+\tilde{q}_{1\lambda}(\eta)
 \end{equation}
 where the $\beta_{\lambda}$ are constants depending on $l$ and $m$ which can be evaluated numerically and the function $f(\eta)$ and the $\beta_{\lambda}$ are chosen such that $\tilde{q}_{1\lambda}$ does not contribute as $\eta\rightarrow 1$. The $\beta_{\lambda}$ coefficients are global in nature and depend on the boundary conditions imposed, as we shall see in Appendix C.
Further following Candelas, we may take
\begin{equation}
f(\eta) = \frac{\sqrt{2}}{32}\int_{1}^{\eta} \frac{\left((\xi+1)^{4}-16\right)}{(\xi^{2}-1)^{3/2}} \rmd\xi,
\end{equation}
so, in particular, have
\begin{equation}
\label{eq:fseries}
f(\eta) = (\eta-1)^{1/2}+\case{1}{32}(\eta-1)^{5/2}+O(\eta-1)^{7/2}\ .
\end{equation}
 A Frobenius analysis on this differential equation about $\eta=1$ now shows that $\tilde{q}_{1\lambda}(\eta)$ satisfies the asymptotic condition
\begin{equation}
\label{eq:qtildeboundary}
\tilde{q}_{1\lambda}(\eta)\sim \case{3}{1792}(\eta-1)^{7/2} \ln(\eta-1) \qquad \mathrm{as}\,\,\eta\rightarrow1,
\end{equation}
which is uniformly valid in $l$ and $m$.
As a consequence, 
Candelas \cite{Candelas:1980zt} has shown that for this choice of $f(\eta)$ and the coefficients $\beta_{\lambda}$, the remainder term $\tilde{q}_{1\lambda}(\eta)$ does not contribute in the limit $\eta\rightarrow 1$;  although Candelas's analysis was for $\lambda $ integer, the argument applies immediately to non-integer $\lambda$. The contribution coming from $\tilde{q}_{1\lambda}(\eta)+\beta_{l}p_{1\lambda}(\eta)$ can now be calculated to leading order in $(\eta-1)$, using the asymptotic forms (\ref{eq:asymp}) for $p_{1\lambda}$ and the fact that $\tilde{q}_{1\lambda}$ does not contribute in the limit, as
\begin{eqnarray}
\fl
\sum_{l=0}^{\infty}\sum_{m=-l}^{l}(2\lambda+1)\frac{\Gamma(\lambda+|m|/\alpha+1)}{\Gamma(\lambda-|m|/\alpha+1)} [P_{\lambda}^{-|m|/\alpha}(\cos\theta)]^{2}[\tilde{q}_{1\lambda}(\eta)+\beta_{\lambda}p_{1\lambda}(\eta)] \nonumber\\
= (\eta-1)^{1/2}A(\cos\theta)+O(\eta-1)^{3/2},
\end{eqnarray}
where
\begin{equation}
A(\cos\theta)=\sum_{l=0}^{\infty}\sum_{m=-l}^{l}(2\lambda+1)\frac{\Gamma(\lambda+|m|/\alpha+1)}{\Gamma(\lambda-|m|/\alpha+1)} [P_{\lambda}^{-|m|/\alpha}(\cos\theta)]^{2}\beta_{\lambda}.
\end{equation}
This sum is highly convergent and easily evaluated numerically. As we show in more detail in Appendix C, the $\beta_{\lambda}$ coefficients are determined by equating the asymptotic expansions for two equivalent representations of the function $q_{1\lambda}(\eta)$, one being an integral representation obtained from the Wronskian condition and the other being the form (\ref{eq:q1def}).
Returning now to the evaluation of $F(\eta,\cos\theta)$,
\begin{eqnarray}
\label{eq:f}
\fl
F(\eta,\cos\theta)=\sum_{l=0}^{\infty}\sum_{m=-l}^{l}(2\lambda+1)\frac{\Gamma(\lambda+|m|/\alpha+1)}{\Gamma(\lambda-|m|/\alpha+1)} [P_{\lambda}^{-|m|/\alpha}(\cos\theta)]^{2}q_{1\lambda}(\eta) \nonumber\\
\fl \qquad\qquad\,\,\,=\sum_{l=0}^{\infty}\sum_{m=-l}^{l}(2\lambda+1)\frac{\Gamma(\lambda+|m|/\alpha+1)}{\Gamma(\lambda-|m|/\alpha+1)} [P_{\lambda}^{-|m|/\alpha}(\cos\theta)]^{2}\Big( \beta_{\lambda}p_{1\lambda}(\eta)-f(\eta)Q_{\lambda}(\eta)\nonumber\\
-\sqrt{2} (\eta^{2}-1)^{1/2} \frac{\rmd Q_{\lambda}(\eta)}{\rmd \eta} +\tilde{q}_{1\lambda}(\eta)\Big).
\end{eqnarray}
For the second and third terms in the expression above, we can use the generalized Heine identity (\ref{eq:generalizedheine}) and its derivative respectively. As a series expansion in $(\eta-1)$, we obtain
\begin{eqnarray}
\label{eq:Fseries}
\fl
F(\eta,\cos\theta) = \frac{2\alpha}{(\eta-1)^{3/2}}-\frac{\alpha}{2(\eta-1)^{1/2}}+(\eta-1)^{1/2}\Big(A(\cos\theta)-\frac{\alpha}{16}+\frac{(\alpha^{2}-1)}{6\alpha\sin^{2}\theta}\nonumber\\
+\frac{(1+10\alpha^{2}-11\alpha^{4})}{90\alpha^{3}\sin^{4}\theta}\Big)+O(\eta-1)^{3/2}.
\end{eqnarray}

We can now calculate the asymptotic series expansions for $g^{\tau\tau'}G_{;\tau\tau'}$, $g^{rr'}G_{;rr'}$ and  $g^{rr'}G_{;r'\theta}$. Employing Eq.(\ref{eq:Fseries}) in expressions (\ref{eq:Gtautaup}), (\ref{eq:Grrp}) and (\ref{eq:Grptheta}) yields
\begin{eqnarray}
\label{eq:Gtautaupseries}
\fl
\qquad [g^{\tau\tau'}G_{;\tau\tau'}]_{\mathrm{r}}=\frac{\kappa^{2}}{8\pi^{2}M^{2}(\eta-1)^{2}}-\frac{\kappa^{2}}{64\pi^{2}M^{2}}+\frac{\kappa^{2}}{96\pi^{2}M^{2}}\frac{(\alpha^{2}-1)}{\alpha^{2}\sin^{2}\theta} \nonumber\\
-\frac{\kappa^{2}}{1440\pi^{2}M^{2}}\frac{(\alpha^{2}-1)(11\alpha^{2}+1)}{\alpha^{4}\sin^{4}\theta}+\frac{\kappa^{2}}{16\pi^{2}M^{2}\alpha} A(\cos\theta) +\Or(\eta-1),
\end{eqnarray}
\begin{eqnarray}
\label{eq:Grrpseries}
\fl
\qquad [g^{r r'}G_{;r r'}]_{\mathrm{r}}=-\frac{3\kappa^{2}}{8\pi^{2}M^{2}(\eta-1)^{2}}+\frac{\kappa^{2}}{8\pi^{2}M^{2}(\eta-1)}-\frac{3\kappa^{2}}{64\pi^{2}M^{2}}+\frac{\kappa^{2}}{96\pi^{2}M^{2}}\frac{(\alpha^{2}-1)}{\alpha^{2}\sin^{2}\theta} \nonumber\\
-\frac{\kappa^{2}}{1440\pi^{2}M^{2}}\frac{(\alpha^{2}-1)(11\alpha^{2}+1)}{\alpha^{4}\sin^{4}\theta}+\frac{\kappa^{2}}{16\pi^{2}M^{2}\alpha} A(\cos\theta) +\Or(\eta-1),
\end{eqnarray}
\begin{equation}
\label{eq:Grpthetaseries}
\fl \qquad [g^{r r'}G_{;r' \theta}]_{\mathrm{r}} = \Or(\eta-1).
\end{equation}


\subsection{Renormalized Stress Tensor}
We have now calculated all the derivatives of the Green's function we require in order to calculate the unrenormalized components of the stress tensor. The term $g^{a b'}G_{;a b'}$ appearing in the stress tensor (\ref{eq:Tunren}) is found by summing the expressions (\ref{eq:Gtautaupseries}), (\ref{eq:Grrpseries}), (\ref{eq:Gthetathetap}) and (\ref{eq:Gphiphip}), yielding
\begin{equation}
[g^{a b'}G_{;a b'}]_{\mathrm{r}}=\frac{\kappa^{2}}{8\pi^{2}M^{2}\alpha}A(\cos\theta)-\frac{\kappa^{2}}{24\pi^{2}M^{2}}\frac{(\alpha^{2}-1)}{\alpha^{2}}\frac{\cos^{2}\theta}{\sin^{4}\theta}+\Or(\eta-1).
\end{equation}
Moreover, the term $\Box G$ appearing in (\ref{eq:Tunren}) vanishes since, by definition, $G(x,x')$ for a massless, Ricci-flat scalar field is a solution to the inhomogeneous wave-equation $\Box G =-\delta$. There is, however, a contribution to the divergent subtraction terms (\ref{eq:tdivhadamard}) coming from $\Box G_{div}$ since this is non-zero. Combining these results, we arrive at the following expression for the unrenormalized stress tensor components near the horizon:
\begin{eqnarray}
\label{eq:tunrentautau}
\fl
\langle \hat{T}^{\tau}{}_{ \tau} \rangle_{unren} =\frac{\kappa^{2}}{8\pi^{2}M^{2}}\frac{1}{(\eta-1)^{2}}-2\xi\frac{\kappa^{2}}{8 \pi^{2}M^{2}}\frac{1}{(\eta-1)}+(14\xi-1)\frac{\kappa^{2}}{64\pi^{2}M^{2}}\nonumber\\
-\frac{\kappa^{2}}{1440\pi^{2}M^{2}}\frac{(\alpha^{2}-1)(11\alpha^{2}+1)}{\alpha^{4}\sin^{4}\theta}\nonumber\\
-(2\xi-\case{1}{2})\frac{\kappa^{2}}{24\pi^{2}M^{2}}\frac{(\alpha^{2}-1)}{\alpha^{2}\sin^{4}\theta} +(\xi-\case{1}{6})\frac{\kappa^{2}}{16\pi^{2}M^{2}}\frac{(\alpha^{2}-1)}{\alpha^{2}\sin^{2}\theta} ,
\end{eqnarray}
\begin{eqnarray}
\label{eq:tunrenrr}
\fl
\langle \hat{T}^{r} {}_{ r}\rangle_{unren} =-\frac{3\kappa^{2}}{8\pi^{2}M^{2}}\frac{1}{(\eta-1)^{2}}+\frac{\kappa^{2}}{8 \pi^{2}M^{2}}\frac{1}{(\eta-1)}+(2\xi-3)\frac{\kappa^{2}}{64\pi^{2}M^{2}}\nonumber\\
+\xi \frac{\kappa^{2}}{8\pi^{2}M^{2}\alpha}A(\cos\theta)-\frac{\kappa^{2}}{1440\pi^{2}M^{2}}\frac{(\alpha^{2}-1)(11\alpha^{2}+1)}{\alpha^{4}\sin^{4}\theta}\nonumber\\
-(2\xi-\case{1}{2})\frac{\kappa^{2}}{24\pi^{2}M^{2}}\frac{(\alpha^{2}-1)}{\alpha^{2}\sin^{4}\theta} +(\xi-\case{1}{6})\frac{\kappa^{2}}{16\pi^{2}M^{2}}\frac{(\alpha^{2}-1)}{\alpha^{2}\sin^{2}\theta} ,
\end{eqnarray}
\begin{eqnarray}
\label{eq:tunrenthetatheta}
\fl
\langle \hat{T}^{\theta} {}_{ \theta}\rangle_{unren}= \frac{\kappa^{2}}{8\pi^{2}M^{2}}\frac{1}{(\eta-1)^{2}}-(1-2\xi)\frac{\kappa^{2}}{16 \pi^{2}M^{2}}\frac{1}{(\eta-1)}+(1-4\xi)\frac{\kappa^{2}}{32\pi^{2}M^{2}}\nonumber\\
+(2\xi-\case{1}{2}) \frac{\kappa^{2}}{8\pi^{2}M^{2}\alpha}A(\cos\theta)-\frac{\kappa^{2}}{1440\pi^{2}M^{2}}\frac{(\alpha^{2}-1)(11\alpha^{2}+1)}{\alpha^{4}\sin^{4}\theta}\nonumber\\
+\xi \frac{\kappa^{2}}{24\pi^{2}M^{2}}\frac{(\alpha^{2}-1)}{\alpha^{2}\sin^{4}\theta}  ,
\end{eqnarray}
\begin{eqnarray}
\label{eq:tunrenphiphi}
\fl
\langle \hat{T}^{\phi} {}_{ \phi}\rangle_{unren}= \frac{\kappa^{2}}{8\pi^{2}M^{2}}\frac{1}{(\eta-1)^{2}}-(1-2\xi)\frac{\kappa^{2}}{16 \pi^{2}M^{2}}\frac{1}{(\eta-1)}+(1-4\xi)\frac{\kappa^{2}}{32\pi^{2}M^{2}}\nonumber\\
+(2\xi-\case{1}{2}) \frac{\kappa^{2}}{8\pi^{2}M^{2}\alpha}A(\cos\theta)+\frac{\kappa^{2}}{480\pi^{2}M^{2}}\frac{(\alpha^{2}-1)(\alpha^{2}+1)}{\alpha^{4}\sin^{4}\theta}\nonumber\\
-(\xi-\case{1}{6}) \frac{\kappa^{2}}{8\pi^{2}M^{2}}\frac{(\alpha^{2}-1)}{\alpha^{2}}\frac{\cos^{2}\theta}{\sin^{4}\theta} ,
\end{eqnarray}
\begin{eqnarray}
\label{eq:tunrenrtheta}
\fl
\langle \hat{T}^{r} {}_{ \theta}\rangle_{unren}=\xi\frac{\kappa^{2}}{360\pi^{2}M}\frac{(\alpha^{2}-1)(11\alpha^{2}+1)\cos\theta}{\alpha^{4}\sin^{5}\theta}+\xi\frac{\kappa^{2}}{48\pi^{2}M}\frac{(\alpha^{2}-1)}{\alpha^{2}}\frac{\cos\theta}{\sin^{3}\theta}.
\end{eqnarray}
Note that the off-diagonal component is regular and vanishing for minimal coupling.

The geometrical subtraction terms are found by obtaining a series expansion for the singular part of the Green's function, then applying the differential operator (\ref{eq:stresstensorop}) and taking partial coincidence limits; this is completely equivalent to, for example, Christensen's method \cite{Christensen:1976vb}. The results are
\begin{eqnarray}
\label{eq:tdivtautau}
\fl 
\qquad \langle \hat{T}^{\tau}{}_{ \tau} \rangle_{div} = \frac{\kappa^{2}}{8\pi^{2}M^{2}}\frac{1}{(\eta-1)^{2}}-2\xi\frac{\kappa^{2}}{8 \pi^{2}M^{2}}\frac{1}{(\eta-1)}+(\xi-\case{1}{16})\frac{\kappa^{2}}{4\pi^{2}M^{2}},
\end{eqnarray}
\begin{eqnarray}
\label{eq:tdivrr}
\fl
\qquad \langle \hat{T}^{r}{}_{ r} \rangle_{div} = -\frac{3\kappa^{2}}{8\pi^{2}M^{2}}\frac{1}{(\eta-1)^{2}}+\frac{\kappa^{2}}{8 \pi^{2}M^{2}}\frac{1}{(\eta-1)}+(\xi-\case{3}{4})\frac{\kappa^{2}}{16\pi^{2}M^{2}},
\end{eqnarray}
\begin{eqnarray}
\label{eq:tdivthetatheta}
\fl
\qquad \langle \hat{T}^{\theta}{}_{ \theta} \rangle_{div} = \frac{\kappa^{2}}{8\pi^{2}M^{2}}\frac{1}{(\eta-1)^{2}}-(1-2\xi)\frac{\kappa^{2}}{16 \pi^{2}M^{2}}\frac{1}{(\eta-1)}-(\xi-\case{7}{20})\frac{\kappa^{2}}{16\pi^{2}M^{2}},
\end{eqnarray}
\begin{eqnarray}
\label{eq:tdivphiphi}
\fl
\qquad \langle \hat{T}^{\phi}{}_{ \phi} \rangle_{div} = \frac{\kappa^{2}}{8\pi^{2}M^{2}}\frac{1}{(\eta-1)^{2}}-(1-2\xi)\frac{\kappa^{2}}{16 \pi^{2}M^{2}}\frac{1}{(\eta-1)}-(\xi-\case{7}{20})\frac{\kappa^{2}}{16\pi^{2}M^{2}},
\end{eqnarray}
\begin{eqnarray}
\label{eq:tdivrtheta}
\fl
\qquad \langle \hat{T}^{r}{}_{ \theta} \rangle_{div} =0.
\end{eqnarray}

Finally, from Eqs.(\ref{eq:tren}), (\ref{eq:tunrentautau})-(\ref{eq:tunrenrtheta}) and (\ref{eq:tdivtautau})-({\ref{eq:tdivrtheta}), we arrive at the renormalized stress tensor for a massless scalar field in the Hartle-Hawking vacuum on the horizon of the Schwarzschild black hole threaded by an infinite cosmic string:
\begin{eqnarray}
\label{eq:trentautau}
\fl
\langle \hat{T}^{\tau}{}_{ \tau} \rangle_{ren} =(\case{2}{15}-\xi)\frac{\kappa^{2}}{32\pi^{2}M^{2}}+\xi \frac{\kappa^{2}}{8\pi^{2}M^{2}\alpha}A(\cos\theta)-\frac{\kappa^{2}}{1440\pi^{2}M^{2}}\frac{(\alpha^{2}-1)(11\alpha^{2}+1)}{\alpha^{4}\sin^{4}\theta}\nonumber\\
-(2\xi-\case{1}{2})\frac{\kappa^{2}}{24\pi^{2}M^{2}}\frac{(\alpha^{2}-1)}{\alpha^{2}\sin^{4}\theta} +(\xi-\case{1}{6})\frac{\kappa^{2}}{16\pi^{2}M^{2}}\frac{(\alpha^{2}-1)}{\alpha^{2}\sin^{2}\theta} ,
\end{eqnarray}
\begin{eqnarray}
\label{eq:trenrr}
\fl 
\langle \hat{T}^{r}{}_{ r} \rangle_{ren} =(\case{2}{15}-\xi)\frac{\kappa^{2}}{32\pi^{2}M^{2}}+\xi \frac{\kappa^{2}}{8\pi^{2}M^{2}\alpha}A(\cos\theta)-\frac{\kappa^{2}}{1440\pi^{2}M^{2}}\frac{(\alpha^{2}-1)(11\alpha^{2}+1)}{\alpha^{4}\sin^{4}\theta}\nonumber\\
-(2\xi-\case{1}{2})\frac{\kappa^{2}}{24\pi^{2}M^{2}}\frac{(\alpha^{2}-1)}{\alpha^{2}\sin^{4}\theta} +(\xi-\case{1}{6})\frac{\kappa^{2}}{16\pi^{2}M^{2}}\frac{(\alpha^{2}-1)}{\alpha^{2}\sin^{2}\theta} ,\end{eqnarray}
\begin{eqnarray}
\label{eq:trenthetatheta}
\fl
\langle \hat{T}^{\theta} {}_{ \theta}\rangle_{ren}= (\case{13}{60}-\xi)\frac{\kappa^{2}}{16\pi^{2}M^{2}}+(2\xi-\case{1}{2}) \frac{\kappa^{2}}{8\pi^{2}M^{2}\alpha}A(\cos\theta)-\frac{\kappa^{2}}{1440\pi^{2}M^{2}}\frac{(\alpha^{2}-1)(11\alpha^{2}+1)}{\alpha^{4}\sin^{4}\theta}\nonumber\\
+\xi \frac{\kappa^{2}}{24\pi^{2}M^{2}}\frac{(\alpha^{2}-1)}{\alpha^{2}\sin^{4}\theta}  ,
\end{eqnarray}
\begin{eqnarray}
\label{eq:trenphiphi}
\fl
\langle \hat{T}^{\phi}{}_{ \phi} \rangle_{ren}=(\case{13}{60}-\xi)\frac{\kappa^{2}}{16\pi^{2}M^{2}}+(2\xi-\case{1}{2}) \frac{\kappa^{2}}{8\pi^{2}M^{2}\alpha}A(\cos\theta)+\frac{\kappa^{2}}{480\pi^{2}M^{2}}\frac{(\alpha^{2}-1)(\alpha^{2}+1)}{\alpha^{4}\sin^{4}\theta}\nonumber\\
-(\xi-\case{1}{6}) \frac{\kappa^{2}}{8\pi^{2}M^{2}}\frac{(\alpha^{2}-1)}{\alpha^{2}}\frac{\cos^{2}\theta}{\sin^{4}\theta} ,
\end{eqnarray}
\begin{eqnarray}
\label{eq:trenrtheta}
\fl
\langle \hat{T}^{r}{}_{ \theta} \rangle_{ren}=\xi\frac{\kappa^{2}}{360\pi^{2}M}\frac{(\alpha^{2}-1)(11\alpha^{2}+1)\cos\theta}{\alpha^{4}\sin^{5}\theta}+\xi\frac{\kappa^{2}}{48\pi^{2}M}\frac{(\alpha^{2}-1)}{\alpha^{2}}\frac{\cos\theta}{\sin^{3}\theta}.
\end{eqnarray}

We note from these expressions that $\langle \hat{T}^{\tau}{}_{ \tau} \rangle_{ren} =\langle \hat{T}^{r}{}_{ r} \rangle_{ren} $, which we would have expected from the symmetries of the black hole horizon. We also see that $\langle \hat{T}^{\theta}{}_{ \theta} \rangle_{ren}\ne \langle \hat{T}^{\phi} {}_{ \phi}\rangle_{ren}$ but that we retrieve equality between these components in the $\alpha\rightarrow 1$ limit. Furthermore, in the conformally coupled case, $\xi=1/6$, we reproduce the Candelas~\cite{Candelas:1980zt} result in the $\alpha\rightarrow 1$ limit. Eqs.(\ref{eq:trentautau})-(\ref{eq:trenphiphi})  also, of course, yield new results for  the renormalized stress tensor on the horizon of a Schwarzschild black hole for general $\xi$, these are trivially obtained from by taking the limit $\alpha\rightarrow 1$.

In Fig.~\ref{fig:trenalpha0.95conf} and Fig.~\ref{fig:trenalpha0.6conf} we have plotted the components of the renormalized stress tensor for the massless, conformally coupled scalar field for $\alpha=0.95$ and $\alpha=0.6$, respectively. We have further plotted the minimally coupled case in Fig.~\ref{fig:trenalpha0.95min} and Fig.~\ref{fig:trenalpha0.6min} for $\alpha=0.95$ and $\alpha=0.6$, respectively. We see that for small azimuthal deficits, the values only move away from the constant Schwarzschild values very close to the poles where the curvature singularity dominates. For this reason, we have included magnified plots Fig.~\ref{fig:trenalpha0.95confzoom} and Fig.~\ref{fig:trenalpha0.95minzoom} for the $\alpha=0.95$ case. For larger deficits ($\alpha=0.6$ for example), we observe that the effects of the string are apparent even relatively close to the equator.

\begin{figure}
\centering
\includegraphics[width=12cm]{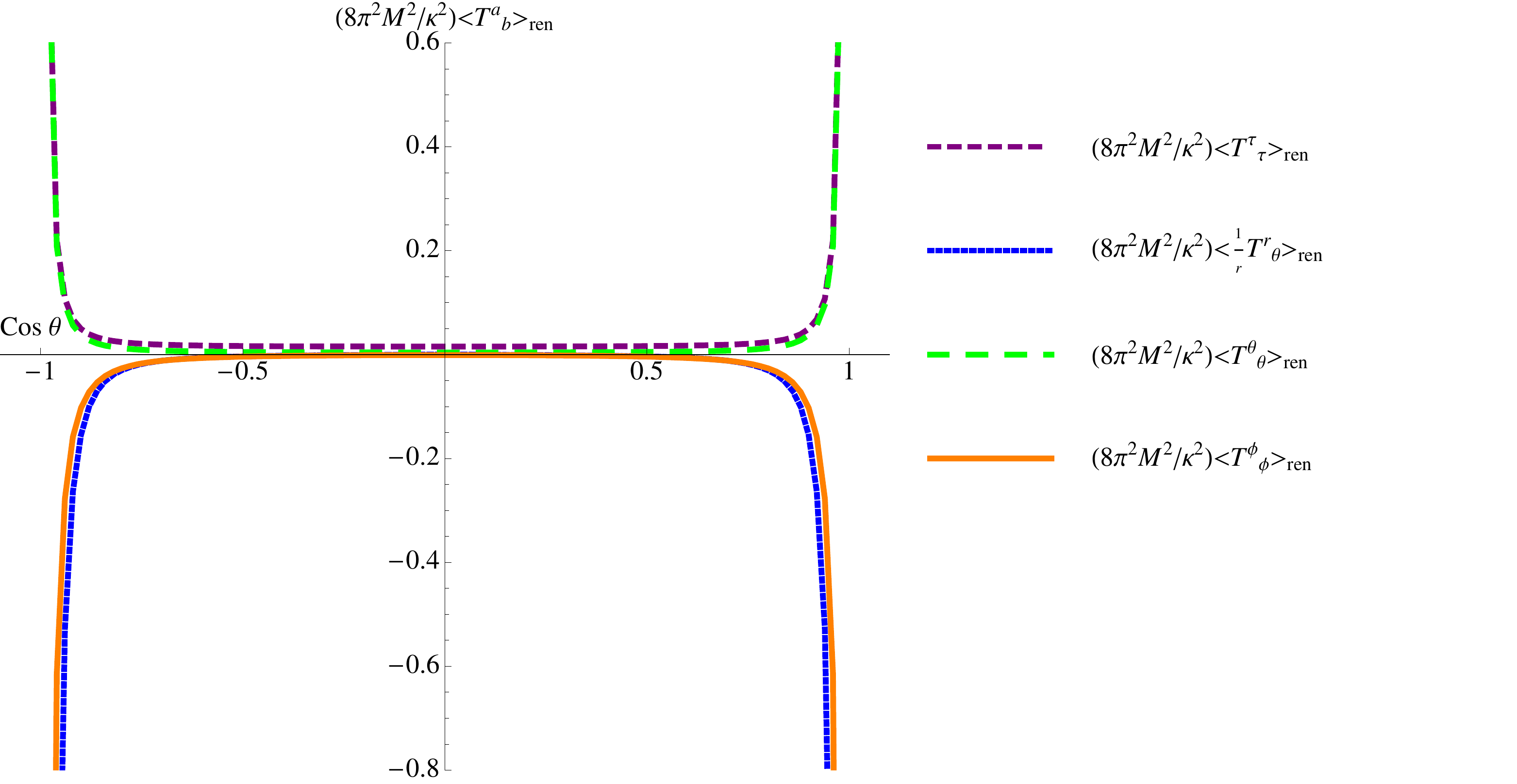}
\caption{\emph{ $(8\pi^{2}M^{2}/\kappa^{2})\langle T^{a}{}_{ b} \rangle_{ren}^{horizon}$} for a conformally coupled scalar field, $\xi=1/6$, as a function of $\cos\theta$ for $\alpha=0.95$.}
\label{fig:trenalpha0.95conf}
\end{figure} 

\begin{figure}
\centering
\includegraphics[width=12cm]{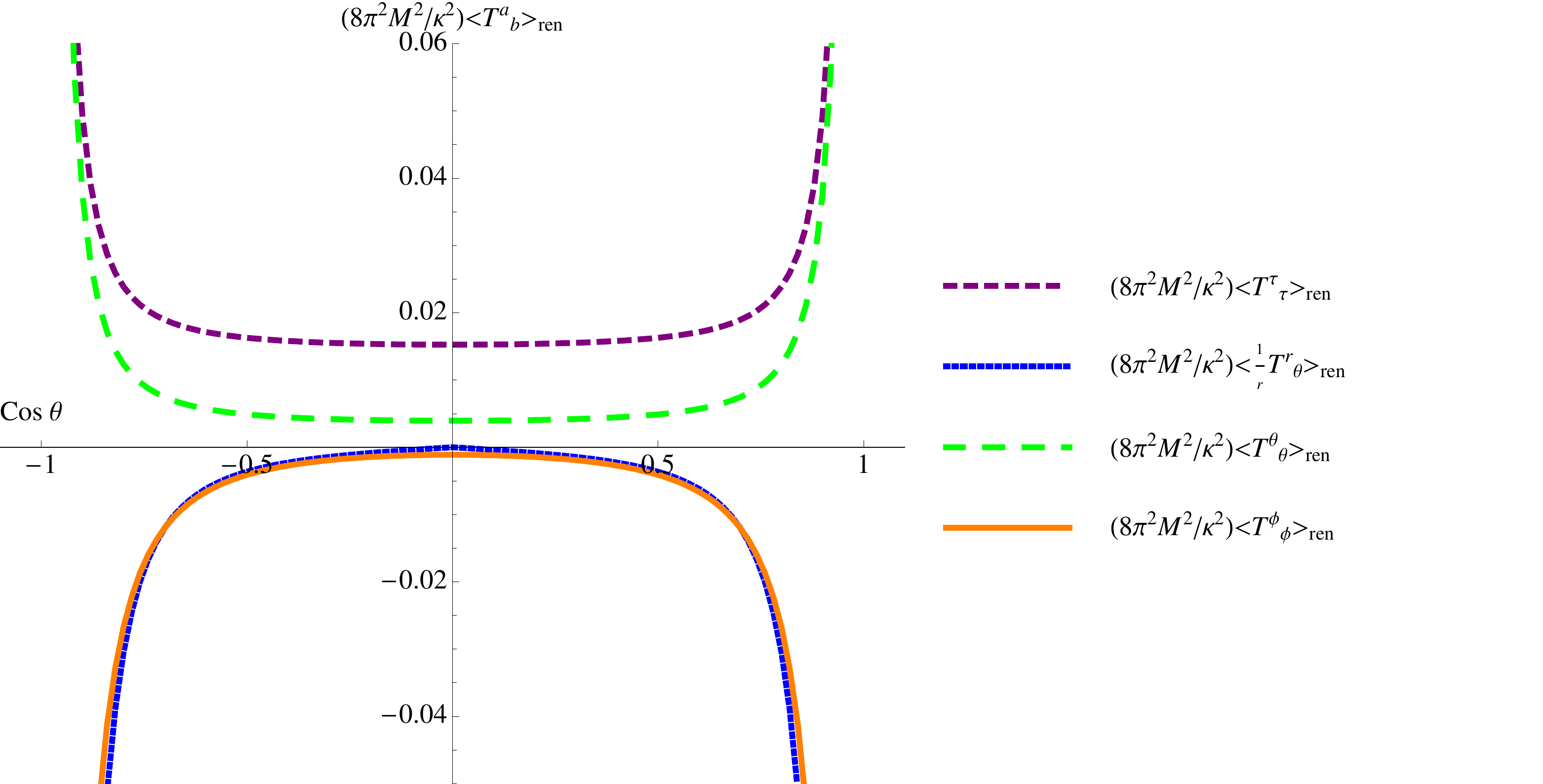}
\caption{\emph{ $(8\pi^{2}M^{2}/\kappa^{2})\langle T^{a}{}_{ b} \rangle_{ren}^{horizon}$} (zoomed in plot) for a conformally coupled scalar field, $\xi=1/6$, as a function of $\cos\theta$ for $\alpha=0.95$.}
\label{fig:trenalpha0.95confzoom}
\end{figure} 

\begin{figure}
\centering
\includegraphics[width=12cm]{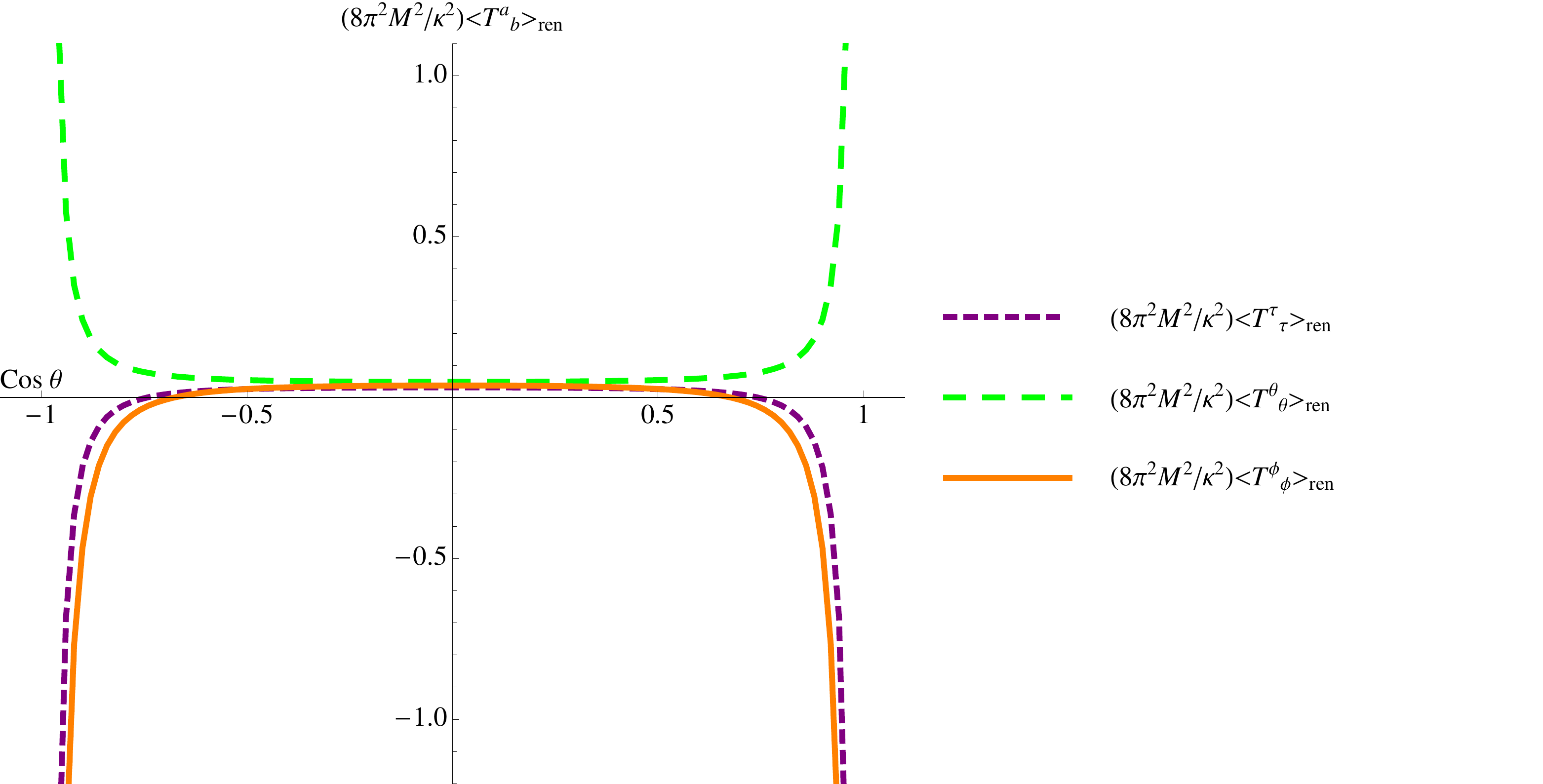}
\caption{\emph{ $(8\pi^{2}M^{2}/\kappa^{2})\langle T^{a}{}_{ b} \rangle_{ren}^{horizon}$} for a minimally coupled scalar field, $\xi=0$, as a function of $\cos\theta$ for $\alpha=0.95$.}
\label{fig:trenalpha0.95min}
\end{figure} 

\begin{figure}
\centering
\includegraphics[width=12cm]{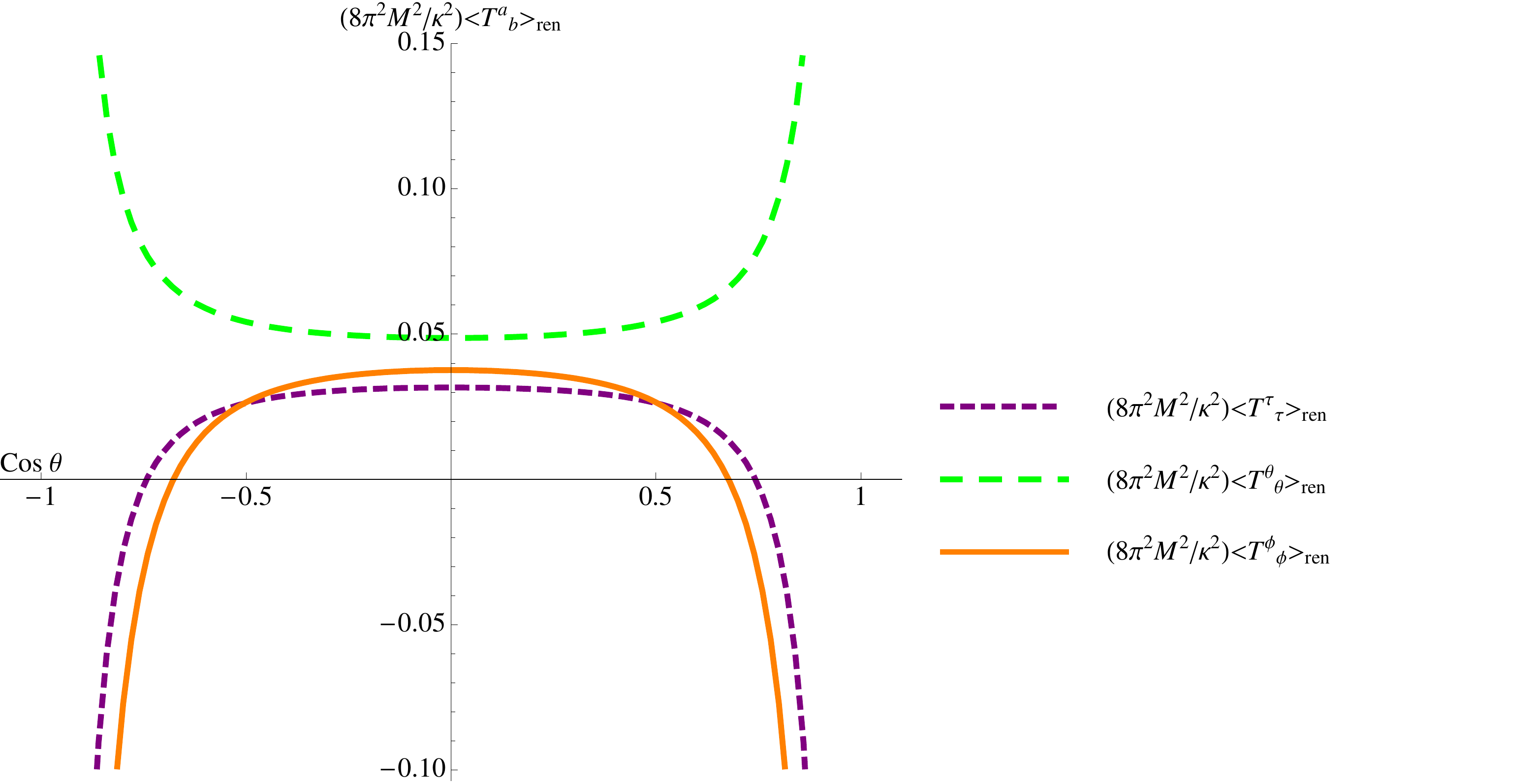}
\caption{\emph{ $(8\pi^{2}M^{2}/\kappa^{2})\langle T^{a}{}_{ b} \rangle_{ren}^{horizon}$} (zoomed in plot) for a minimally coupled scalar field, $\xi=0$, as a function of $\cos\theta$ for $\alpha=0.95$.}
\label{fig:trenalpha0.95minzoom}
\end{figure} 

\begin{figure}
\centering
\includegraphics[width=12cm]{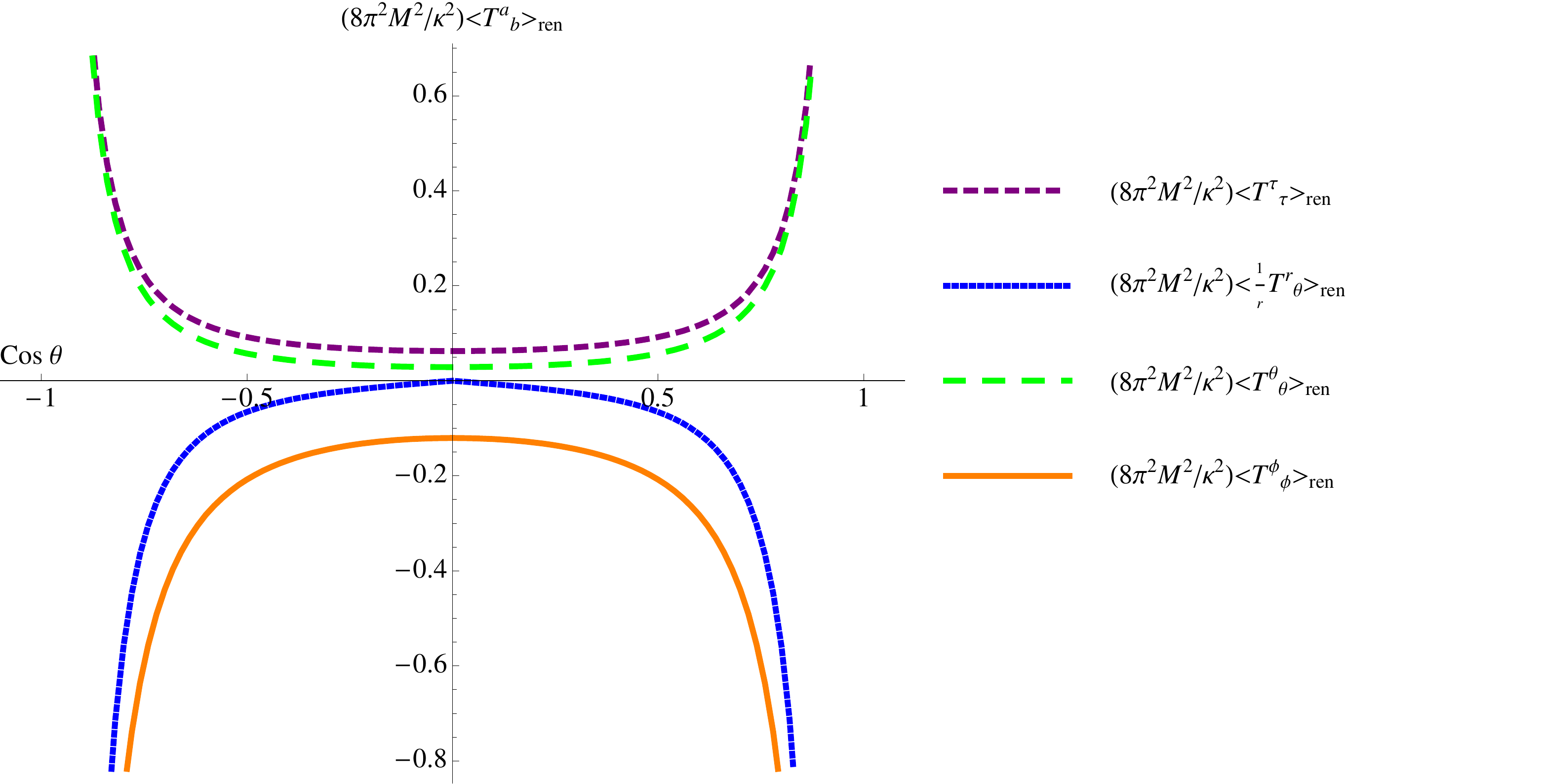}
\caption{\emph{ $(8\pi^{2}M^{2}/\kappa^{2})\langle T^{a}{}_{ b} \rangle_{ren}^{horizon}$} for a conformally coupled scalar field, $\xi=1/6$, as a function of $\cos\theta$ for $\alpha=0.6$.}
\label{fig:trenalpha0.6conf}
\end{figure} 

\begin{figure}
\centering
\includegraphics[width=12cm]{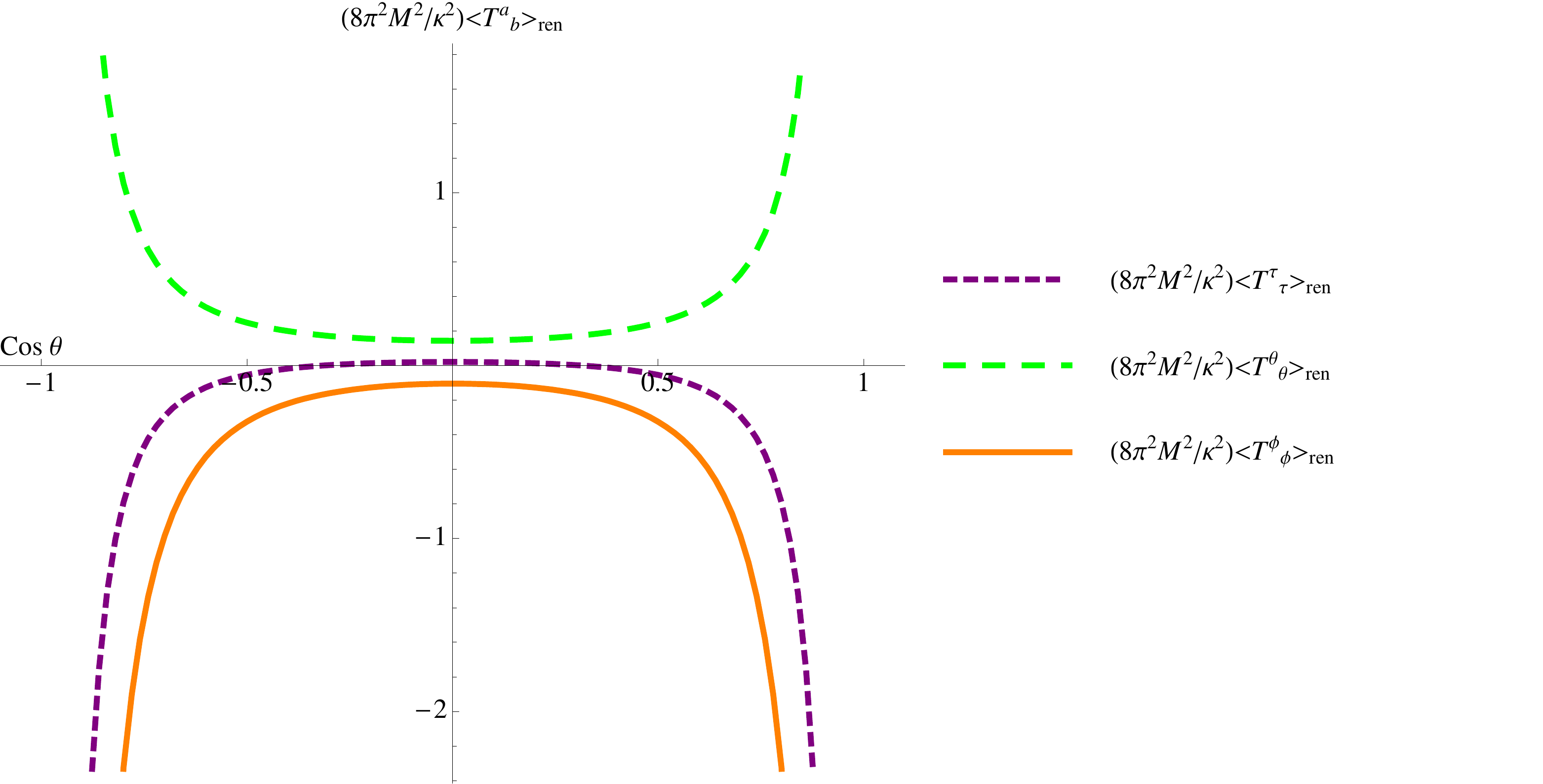}
\caption{\emph{ $(8\pi^{2}M^{2}/\kappa^{2})\langle T^{a}{}_{ b} \rangle_{ren}^{horizon}$} for a minimally coupled scalar field, $\xi=0$, as a function of $\cos\theta$ for $\alpha=0.6$.}
\label{fig:trenalpha0.6min}
\end{figure}


\section{The Conservation Equations: The Minimally Coupled Case}
\label{sec:conservation}
In this section, we give an example of how the stress tensor we have calculated may be shown to satisfy the conservation equations. For simplicity, we consider the minimally coupled case $\xi=0$. The conservation equations are
\begin{equation}
\nabla_{a} \langle\hat{T}^{a}{}_{b}\rangle_{ren} =\partial_{a}\langle\hat{T}^{a}{}_{b}\rangle_{ren}+\Gamma^{a}_{ac}\langle\hat{T}^{c}{}_{b}\rangle_{ren}-\Gamma^{c}_{ab}\langle\hat{T}^{a}{}_{c}\rangle_{ren}=0.
\end{equation}
In particular taking $b=\theta$  we have
\begin{eqnarray}
\label{eq:thetaconservation}
\nabla_{a}\langle\hat{T}^{a}{}_{\theta}\rangle_{ren} = &\partial_{\theta}\langle\hat{T}^{\theta}{}_{\theta}\rangle_{ren}+\partial_{r}\langle\hat{T}^{r}{}_{\theta}\rangle_{ren}+\frac{1}{r}\langle\hat{T}^{r}{}_{\theta}\rangle_{ren}+(r-2 M)\langle\hat{T}^{\theta}{}_{r}\rangle_{ren}+\nonumber\\
&\cot\theta \langle\hat{T}^{\theta}{}_{\theta}\rangle_{ren}-\cot\theta \langle\hat{T}^{\phi}{}_{\phi}\rangle_{ren} =0 .
\end{eqnarray}
Now
\begin{eqnarray}
(r-2M)\langle\hat{T}^{\theta}{}_{r}\rangle_{ren} = \frac{1}{r}\langle\hat{T}^{r}{}_{\theta}\rangle_{ren}\rightarrow 0
 \qquad \mathrm{as}\,\,r\rightarrow 2M
\end{eqnarray}
for the minimally coupled case. However, the derivative $\partial_{r}\langle\hat{T}^{r}{}_{\theta}\rangle_{ren}$ is non-zero on the horizon, as we shall see below. 
For $\xi=0$, the classical expression for this term is given by
\begin{equation}
\label{eq:diffTrtheta}
\partial_{r}\langle\hat{T}^{r}{}_{\theta}\rangle_{ren}=\partial_{r}(g^{rr}\varphi_{,r})\varphi_{,\theta}+g^{rr'}\varphi_{,r}\varphi_{,r\theta}
\end{equation}
and in the regularized version terms like $\varphi_{,r}\varphi_{,\theta}$ are reexpressed with a radial derivative of the Green's function at one spacetime point and a $\theta$ derivative at the other spacetime point. We consider the first term on the right-hand side of Eq.(\ref{eq:diffTrtheta}), taking the radial derivatives with respect to the outer point and taking the inner point to lie on the horizon, then only the $n=0$ term survives and we can differentiate our Heine identity (\ref{eq:generalizedheine}) directly,
\begin{equation}
\fl
\partial_{r}(g^{rr}\varphi_{,r})\varphi_{,\theta}=\frac{\kappa^{2}}{2\pi^{2}M^{2}\alpha} \frac{\partial}{\partial \eta} \left(\frac{(\eta-1)}{(\eta+1)}\frac{\partial^{2}}{\partial\eta\partial\theta'}\left(\frac{\sinh(\chi/\alpha)}{\sin\theta\sin\theta'\sinh\chi(\cosh(\chi/\alpha)-1)}\right)\right)
\end{equation}
where
\begin{equation}
\cosh\chi=\frac{\eta-\cos\theta\cos\theta'}{\sin\theta\sin\theta'} \ ,
\end{equation}
and we have taken $\tau\rightarrow\tau'$ and $\phi\rightarrow\phi'$. Finally, performing the differentiation and taking $\theta\rightarrow\theta'$, we have
\begin{equation}
[\partial_{r}(g^{rr}\varphi_{,r})\varphi_{,\theta}]_\mathrm{r}=- \frac{\kappa^{2}}{360\pi^{2}M^{2}}\frac{(\alpha^2-1)(11\alpha^{2}+1)\cos\theta}{\alpha^{4}\sin^{4}\theta}+O(\eta-1)^{1/2}.
\end{equation}
In taking the $\theta\rightarrow\theta'$ limit, we have made use of Eq.(\ref{eq:thetalimit}).

For the second term on the right-hand side of Eq.(\ref{eq:diffTrtheta}), we have
\begin{eqnarray}
\fl
g^{rr'}\varphi_{,r}\varphi_{,r\theta}=(1-2M/r)^{1/2}(1-2M/r')^{1/2}\frac{\partial}{\partial r}\frac{\partial}{\partial \theta}\frac{\partial}{\partial r'}\Big(\frac{\kappa}{8\pi^{2}}\sum_{n=-\infty}^{\infty}\sum_{m=-\infty}^{\infty}\sum_{l=|m|}^{\infty}(2\lambda+1)  \nonumber\\
\times \frac{\Gamma(\lambda+|m|/\alpha+1)}{\Gamma(\lambda-|m|/\alpha+1)} P^{-|m|/\alpha}_{\lambda}(\cos\theta)P^{-|m|/\alpha}_{\lambda}(\cos\theta')\chi_{n\lambda}(r,r')\Big)
\end{eqnarray}
where we have taken the partial limits $\phi\rightarrow\phi'$, $\tau\rightarrow\tau'$. Taking the `primed' coordinates to correspond the inner spacetime point, we can see from the asymptotic forms (\ref{eq:asymp}) that taking $r'$ to lie on the horizon eliminates all but the $n=\pm1$ modes. The contribution coming from these modes is
\begin{equation}
\frac{\kappa^{2}}{8\sqrt{2}\,\pi^{2}M^{2}\alpha}\frac{(\eta^{2}-1)^{1/2}}{(\eta+1)}\frac{\partial}{\partial \eta}\frac{\partial}{\partial \theta} F(\eta,\cos\theta)
\end{equation}
where $F(\eta,\cos\theta)$ is given by Eq.(\ref{eq:f}) and in going to the limit $\theta\rightarrow\theta'$ we have again made use of Eq.(\ref{eq:thetalimit}). Now, using our series solution  (\ref{eq:Fseries}) for $F(\eta,\cos\theta)$, we arrive at
\begin{eqnarray}
\fl
[g^{rr'}\varphi_{,r}\varphi_{,r\theta}]_\mathrm{r}=\frac{\kappa^{2}}{16\pi^{2}M^{2}\alpha}\frac{\partial}{\partial\theta}A(\cos\theta)+\frac{\kappa^{2}}{360\pi^{2}M^{2}}\frac{(\alpha^{2}-1)(11\alpha^{2}+1)\cos\theta}{\alpha^{4}\sin^{5}\theta} \nonumber\\
-\frac{\kappa^{2}}{48\pi^{2}M^{2}}\frac{(\alpha^{2}-1)}{\alpha^{2}}\frac{\cos\theta}{\sin^{3}\theta}+O(\eta-1)
\end{eqnarray}
Combining these results, we have
\begin{equation}
\label{eq:diffTrthetaren}
[\partial_{r}\langle \hat{T}^{r}{}_{\theta}\rangle_{ren}]=\frac{\kappa^{2}}{16\pi^{2}M^{2}\alpha}\frac{\partial}{\partial\theta}A(\cos\theta)-\frac{\kappa^{2}}{48\pi^{2}M^{2}}\frac{(\alpha^{2}-1)}{\alpha^{2}}\frac{\cos\theta}{\sin^{3}\theta}.
\end{equation}
There are no subtraction terms for this quantity so that this expression represents the renormalized expectation value $\partial_{r}\langle\hat{T}^{r}_{\theta}\rangle_{ren}$. Substituting Eqs.(\ref{eq:diffTrthetaren}), (\ref{eq:trenthetatheta}) and (\ref{eq:trenphiphi}), for $\xi=0$, into Eq.(\ref{eq:thetaconservation}), it is readily seen that
\begin{equation}
\nabla_{a}\langle\hat{T}^{a}{}_{\theta}\rangle_{ren}=0
\end{equation}
as required. One can similarly, show that $\nabla_{a}\langle\hat{T}^{a}{}_{r}\rangle_{ren}=0$. The equations for $\tau$ and $\phi$ are trivially satisfied.


\section{Conclusions}
We have considered  a massless, arbitrarily coupled quantum scalar field on a Schwarzschild black hole threaded by an infinite straight cosmic string. We have obtained simple expressions for the both the renormalized vacuum polarization and the renormalized stress tensor on the black hole horizon for this field in the Hartle-Hawking vacuum state, which is a two-fold generalization of the Candelas \cite{Candelas:1980zt} result to encompass an arbitrary coupled field and a black hole threaded by a cosmic string. In order to do this, we have derived a very useful summation formula involving non-integer Legendre functions, which we have called a generalized Heine Identity given its analogy with the standard Heine Identity involving Legendre functions of integer degree. The derivation of this formula involved equating equivalent expressions for the same Green's function on a conveniently chosen space-time. Furthermore, the derivation did not rely on specific properties of the Legendre functions. This idea can be used to derive many useful summation formulae and addition theorems for non-integer Legendre functions, some of which are presented in Appendix B.  Elsewhere \cite{CSHorizon}, we have used this idea to obtain an expression for the vacuum polarization on the entire exterior region of the space-time. We believe this framework will prove useful in the other axially symmetric black hole calculations, most importantly, the Kerr-Newman black hole. 


\appendix
\section*{Appendix A: Normalization Condition}
\setcounter{section}{1}
In this Appendix, we shall prove the normalization condition Eq.(\ref{eq:norm}) for the non-integer Legendre functions that arise on separation of the wave-equation on cosmic string spacetimes. We begin by noting the following relationship between the Legendre function and the Gegenbauer polynomial \cite{Erdelyi},
\begin{equation}
\label{eq:gegenbauer}
P_{n+\nu-1/2}^{1/2-\nu}(x)=\frac{2^{1/2-\nu}\Gamma(2\nu)\Gamma(n+1)}{\Gamma(n+2\nu)\Gamma(\nu+1/2)}(1-x^{2})^{\nu/2-1/4} C_{n}^{\nu}(x)
\end{equation}
where $C_{n}^{\nu}(z)$ is the Gegenbauer polynomial and $n$ is an integer. We also have a normalization for these polynomials \cite{Erdelyi} given by
\begin{equation}
\label{eq:normgegen}
\int_{-1}^{1}C_{n}^{\nu}(x) C_{n'}^{\nu}(x) (1-x^{2})^{\nu-1/2} dx=\frac{2^{1-2\nu} \pi \Gamma(n+2\nu)}{(n+\nu)\Gamma(n+1)\Gamma(\nu)^{2}} \delta_{n n'}.
\end{equation}
Choosing $n=l-|m|$ and $\nu=1/2+|m|/\alpha$ in Eq.(\ref{eq:gegenbauer}) gives us the appropriate Legendre function in terms of the Gegenbauer polynomials, and with this choice of $n$ and $\nu$, Eq.(\ref{eq:normgegen}) becomes
\begin{eqnarray}
\label{eq:normproof}
\int_{-1}^{1}P_{l-|m|+|m|/\alpha}^{-|m|/\alpha}(\cos\theta) P_{l'-|m|+|m|/\alpha}^{-|m|/\alpha}(\cos\theta) \rmd (\cos\theta)= \nonumber\\
\frac{2^{-4|m|/\alpha} \pi \Gamma(1+2|m|/\alpha)^{2}}{\Gamma(1+|m|/\alpha)^{2}\Gamma(1/2+|m|/\alpha)^{2}}\frac{2}{(2\lambda+1)}\frac{\Gamma(\lambda-|m|/\alpha+1)}{\Gamma(\lambda+|m|/\alpha+1)} \delta_{l l'}.
\end{eqnarray}
We now use the so-called doubling formula for the Gamma function \cite{gradriz}
\begin{equation}
\Gamma(1+2|m|/\alpha)=\frac{2^{2|m|/\alpha}}{\sqrt{\pi}}\Gamma(1/2+|m|/\alpha)\Gamma(1+|m|/\alpha)
\end{equation}
which implies that
\begin{equation}
\frac{2^{-4|m|/\alpha} \pi \Gamma(1+2|m|/\alpha)^{2}}{\Gamma(1+|m|/\alpha)^{2}\Gamma(1/2+|m|/\alpha)^{2}}=1.
\end{equation}
Finally, substituting this into Eq.(\ref{eq:normproof}) gives the desired result, \textit{viz.},
\begin{eqnarray}
&\int_{-1}^{1}P_{l-|m|+|m|/\alpha}^{-|m|/\alpha}(\cos\theta) P_{l'-|m|+|m|/\alpha}^{-|m|/\alpha}(\cos\theta) d(\cos\theta) 
=\nonumber\\
&\qquad\qquad\qquad\frac{2}{(2\lambda+1)}\frac{\Gamma(\lambda-|m|/\alpha+1)}{\Gamma(\lambda+|m|/\alpha+1)}\delta_{ll'}. 
\end{eqnarray}


\section*{Appendix B: Summation Formulae}
\setcounter{section}{2}
In this Appendix, we shall derive some interesting summation formulae for non-integer Legendre functions based on considering equivalent forms of the 3D Green's function on flat space threaded by a cosmic string
\begin{equation}
\rmd s^{2}=\rmd r^{2}+r^{2} \rmd\theta^{2} + \alpha^{2} r^{2}\sin^{2}\theta \rmd\phi^{2}.
\end{equation}
The 3D Green's function for Laplace's equation satisfying
\begin{equation}
\nabla^2 G_3(\mathbf{x},\mathbf{x'}) = -{g_3}(\mathbf{x})^{-1/2} \delta(\mathbf{x}-\mathbf{x'})
\end{equation}
can be written in spherical polar coordinates as the mode sum:
\begin{eqnarray}
\label{eq:app1}
\fl
G_3(\mathbf{x},\mathbf{x'})=\frac{1}{4\pi\alpha}&\sum_{m=-\infty}^{\infty} \rme^{im (\phi-\phi')}\sum_{l=|m|}^{\infty}\frac{\Gamma(\lambda+|m|/\alpha+1)}{\Gamma(\lambda-|m|/\alpha+1)}\nonumber\\ &\qquad P^{-|m|/\alpha}_{\lambda}(\cos\theta)P^{-|m|/\alpha}_{\lambda}(\cos\theta')(r_{<})^{\lambda} (r_{>})^{-(\lambda+1)} .
\end{eqnarray}
Note that this corresponds to the $n=0$ term of the 4D Green function in the corresponding ultrastatic ($g_{tt}=-1$) space-time, but that the
corresponding statement is not true in the merely static Schwarzschild cosmic string space-time.
We can also write this same Green's function in several other forms and in several coordinate systems, for example,
performing the mode decomposition in cylindrical polar coordinates, $\rho=r \sin\theta$ and $z=r\cos\theta$, we have
\begin{eqnarray}
\label{eq:app2}
\fl
G_3(\mathbf{x},\mathbf{x'})=\frac{1}{2 \pi^{2} \alpha} \sum_{m=-\infty}^{\infty} \rme^{im(\phi-\phi')}\int_{0}^{\infty} \cos k(z-z') I_{|m|/\alpha}(k\rho_{<}) K_{|m|/\alpha}(k\rho_{>}) \rmd k  .
\end{eqnarray}
The integral over $k$ may further be performed to yield
\begin{eqnarray}
\label{eq:app3}
\fl
G_3(\mathbf{x},\mathbf{x'})=\frac{1}{4 \pi^{2}\alpha (\rho \rho')^{1/2}} \sum_{m=-\infty}^{\infty} e^{im (\phi-\phi')} Q_{|m|/\alpha-1/2}\left(\frac{(z-z')^{2}+\rho^{2}+\rho'^{2}}{2 \rho \rho'}\right) .
\end{eqnarray}
Another form of this Green's function is obtained from General Axi-Symmetric Potential Theory~\cite{Weinstein, Linet1977}:
\begin{eqnarray}
\label{eq:app4}
\fl
G_3(\mathbf{x},\mathbf{x'})=\frac{1}{4 \pi^{2} \alpha}& \sum_{m=-\infty}^{\infty}\rme^{im(\phi-\phi')}\nonumber\\
 & \int_{0}^{\pi} \frac{(\rho \rho')^{|m|/\alpha} \sin^{2|m|/\alpha} \Psi }{[(z-z')^{2}+\rho^{2}+\rho'^{2}-2 \rho \rho' \cos\Psi]^{(|m|/\alpha+1/2)}} \rmd\Psi.
\end{eqnarray}
These expressions are valid for all $\alpha \in \mathbb{R}_+$; equating any one of these expressions with another gives us a formula for the functions involved. Some of these results are well known and can be found in volumes such as  Gradsteyn \& Rhyzik~\cite{gradriz}, nevertheless, we have non-obvious, important results, for example,  equating (\ref{eq:app1}) with (\ref{eq:app3}) and taking $r\rightarrow r'$, we obtain
\begin{eqnarray}
\label{eq:app5}
\fl \sum_{l=|m|}^{\infty}
\frac{\Gamma(\lambda+|m|/\alpha+1)}{\Gamma(\lambda-|m|/\alpha+1)}&& P^{-|m|/\alpha}_{\lambda}(\cos\theta)P^{-|m|/\alpha}_{\lambda}(\cos\theta')=\nonumber\\
&&
\frac{1}{\pi (\sin\theta\sin\theta')^{1/2}} Q_{|m|/\alpha-1/2}\left(\frac{1-\cos\theta\cos\theta'}{\sin\theta\sin\theta'}\right).
\end{eqnarray}
Also, equating Eq.(\ref{eq:app1}) with Eq.(\ref{eq:app4}) we arrive at the following formula
\begin{eqnarray}
\fl \sum_{l=|m|}^{\infty}
\frac{\Gamma(\lambda+|m|/\alpha+1)}{\Gamma(\lambda-|m|/\alpha+1)}&& P^{-|m|/\alpha}_{\lambda}(\cos\theta)P^{-|m|/\alpha}_{\lambda}(\cos\theta')(r_{<})^{\lambda} (r_{>})^{-(\lambda+1)} =\nonumber\\
&&
\frac{1}{\pi} \int_{0}^{\pi} \frac{(\rho \rho')^{|m|/\alpha} \sin^{2|m|/\alpha} \Psi }{[(z-z')^{2}+\rho^{2}+\rho'^{2}-2 \rho \rho' \cos\Psi]^{(|m|/\alpha+1/2)}} \rmd\Psi ,
\end{eqnarray}
so that the integral expression on the RHS is a type of generating function for these non-integer Legendre functions.

There are also a set of formulae that hold for a more restrictive set of $\alpha$ values. Linet \cite{LinetCosmicString1987} has shown that, when $\alpha > 1/2$, the Green's function may be written as
\begin{equation}
G_3(\mathbf{x},\mathbf{x'}) = \frac{1}{4 \pi \sigma}+\frac{1}{8\pi^{2}\alpha} \int_{0}^{\infty} \frac{1}{R} F_{\alpha}(u,\phi-\phi') \rmd u
\end{equation}
where
\begin{eqnarray}
\sigma&=[\rho^{2}+\rho'^{2}+(z-z')^{2}-2\rho\rho' \cos\alpha(\phi-\phi')]^{1/2} \nonumber\\
R&=[\rho^{2}+\rho'^{2}+(z-z')^{2}+2\rho\rho'\cosh u]^{1/2}
\end{eqnarray}
and
\begin{eqnarray}
\fl
F_{\alpha}(u,\Psi) = \frac{\sin(\Psi-\pi/\alpha)}{\cosh(u/\alpha)-\cos(\Psi-\pi/\alpha)}-\frac{\sin(\Psi+\pi/\alpha)}{\cosh(u/\alpha)-\cos(\Psi+\pi/\alpha)}.
\end{eqnarray}
Equating this to (\ref{eq:app1}) for $\alpha>1/2$ and taking $r\rightarrow r'$, we find
\begin{eqnarray}
\fl
& \frac{1}{\alpha}\sum_{m=-\infty}^{\infty}e^{im (\phi-\phi')}\sum_{l=|m|}^{\infty}\frac{\Gamma(\lambda+|m|/\alpha+1)}{\Gamma(\lambda-|m|/\alpha+1)} P^{-|m|/\alpha}_{\lambda}(\cos\theta)P^{-|m|/\alpha}_{\lambda}(\cos\theta')\nonumber\\
\fl &\qquad =\frac{1}{ [2(1-\cos\alpha\gamma)]^{1/2}}+\frac{1}{2\pi\alpha}\frac{1}{(2 \sin\theta \sin\theta')^{1/2}} \int_{0}^{\infty}\frac{F_{\alpha}(u,\phi-\phi')}{[\cosh\xi+\cosh u]^{1/2}} \rmd u
\end{eqnarray}
where
\begin{eqnarray}
\cos\alpha\gamma&= \cos\theta\cos\theta'+\sin\theta\sin\theta'\cos\alpha(\phi-\phi') \nonumber\\
\cosh\xi &= \frac{1-\cos\theta\cos\theta'}{\sin\theta\sin\theta'}.
\end{eqnarray}
This formula has the benefit of relating a 3D mode-sum expression to an expression that isolates the 3D Hadamard singularity structure
(the first term on the RHS) together with a regular term that depends on the boundary conditions. We have used this result to prove the regularity of the mode-sum that arises in the calculation of the vacuum polarization on the exterior region of the spacetime \cite{CSHorizon}. 

Another form of the Green's function can be given in terms of the half-integer Legendre functions by considering the Green's function in toroidal coordinates, which are related to the cartesian coordinates by
\begin{eqnarray}
\fl x=\frac{\sinh\mu \cos\phi}{\cosh\mu-\cos\eta} \qquad y=\frac{\sinh\mu\sin\phi}{\cosh\mu-\cos\eta}\qquad  z=\frac{\sin\eta}{\cosh\mu-\cos\eta},
 \end{eqnarray}
 where the ranges of the coordinates are
$ 0\le\mu <\infty$, $0\le\eta < 2 \pi$,  $0\le\phi < 2 \pi$.
The mode-sum form of the Green's function obtained by separating in these coordinates is
\begin{eqnarray}
\fl
&G_3(\mathbf{x},\mathbf{x'})= \frac{1}{4\pi^2 \alpha}\sum_{m=-\infty}^{\infty}\rme^{im(\phi-\phi')}\sum_{n=-\infty}^{\infty}\rme^{i n(\eta-\eta')}[(\cosh\mu-\cos\eta)(\cosh\mu'-\cos\eta')]^{1/2} \nonumber\\
\fl&\qquad
   \rme^{i|m|\pi/\alpha} \frac{\Gamma(n+|m|/\alpha+1/2)}{\Gamma(n-|m|/\alpha+1/2)}P_{n-1/2}^{-|m|/\alpha}(\cosh\mu_{<}) Q_{n-1/2}^{-|m|/\alpha}(\cosh\mu_{>}) .
\end{eqnarray}
Equating this to (\ref{eq:app3}) gives us the following addition theorem
\begin{eqnarray}
\fl
&\sum_{n=-\infty}^{\infty}\rme^{i n(\eta-\eta')}\rme^{i|m|\pi/\alpha}\frac{\Gamma(n+|m|/\alpha+1/2)}{\Gamma(n-|m|/\alpha+1/2)}P_{n-1/2}^{-|m|/\alpha}(\cosh\mu_{<}) Q_{n-1/2}^{-|m|/\alpha}(\cosh\mu_{>}) \nonumber\\
\fl&\qquad
=\frac{1}{(\sinh \mu \sinh \mu')^{1/2}}Q_{|m|/\alpha-1/2}(\chi).
\end{eqnarray}
where
\begin{equation}
\chi=\frac{(z-z')^{2} + \rho^{2}+\rho'^{2}}{2\rho\rho'}
\end{equation}
which are related to the toroidal coordinates by
\begin{equation}
\rho=\frac{\sinh\mu}{\cosh\mu-\cos\eta} \qquad\qquad z=\frac{\sin\eta}{\cosh\mu-\cos\eta}.
\end{equation}

The final coordinate system we shall investigate is prolate spheroidal coordinates which may trivially extended to the oblate spheroidal case. The relationship between prolate spheroidal coordinates and cartesian coordinates is
given by
\begin{equation}
\fl
x=\sinh\sigma\sin\theta\cos\phi \qquad y=\sinh\sigma\sin\theta\sin\phi  \qquad z=\cosh\sigma\cos\theta.
\end{equation}
where the ranges of the coordinates are $ 0\le\sigma <\infty$, $0\le\theta\le \pi$,  $0\le\phi
< 2\pi$.
The mode form of the Green's function obtained by separating in these coordinates is
\begin{eqnarray}
\fl
& G_3(\mathbf{x},\mathbf{x'}) = \frac{1}{4\pi\alpha} \sum_{m=-\infty}^{\infty} \rme^{i m (\phi-\phi')}\sum_{l=|m|}^{\infty}(2\lambda+1)\rme^{i|m|\pi/\alpha}\frac{\Gamma(\lambda+|m|/\alpha+1)^{2}}{\Gamma(\lambda-|m|/\alpha+1)^{2}}  \nonumber\\
\fl & \qquad P_{\lambda}^{-|m|/\alpha}(\cos\theta)P_{\lambda}^{-|m|/\alpha}(\cos\theta')P_{\lambda}^{-|m|/\alpha}(\cosh\sigma_{<})Q_{\lambda}^{-|m|/\alpha}(\cosh\sigma_{>})
\end{eqnarray}
We shall write down only one possible summation formula here, by equating the Green's function expression above to (\ref{eq:app3}), we obtain a summation formula for a product of four Legendre functions:
\begin{eqnarray}
\fl &
\sum_{l=|m|}^{\infty}(2\lambda+1)\rme^{i|m|\pi/\alpha}\frac{\Gamma(\lambda+|m|/\alpha+1)^{2}}{\Gamma(\lambda-|m|/\alpha+1)^{2}}  P_{\lambda}^{-|m|/\alpha}(\cos\theta)P_{\lambda}^{-|m|/\alpha}(\cos\theta')  \nonumber\\
\fl & \qquad P_{\lambda}^{-|m|/\alpha}(\cosh\sigma_{<})Q_{\lambda}^{-|m|/\alpha}(\cosh\sigma_{>})  
=\frac{1}{\pi} \frac{Q_{|m|/\alpha-1/2}(\chi)}{\sqrt{\sinh\sigma\sinh\sigma'\sin\theta\sin\theta'}}
\end{eqnarray}
where
\begin{equation}
\fl
\chi=\frac{\cosh^2\sigma+\cosh^2\sigma'-\sin^2\theta-\sin^2\theta'-2\cosh\sigma\cosh\sigma'\cos\theta\cos\theta'}{2\sinh\sigma\sinh\sigma'\sin\theta\sin\theta'}
\end{equation}
There are many more summation and addition formulae that can be derived in this way. Those presented here were either of particular use to us elsewhere or are to the best of our knowledge previously unpublished.


\section*{Appendix C: Determining the $\beta_{\lambda}$ Coefficients}
\setcounter{section}{3}
\label{sec:beta}
 
We briefly describe the procedure for determining the $\beta_{\lambda}$ coefficients occurring in the definition of $A(\cos\theta)$. The method described is completely analogous to the Candelas \cite{Candelas:1980zt} method except that we replace $l$ by $\lambda=l-|m|+|m|/\alpha$ everywhere it occurs. We begin with the Wronskian relation between the $n=\pm1$ radial functions of the first and second kind,
\begin{equation}
p_{1\lambda}(\eta)\frac{\rmd}{\rmd\eta}q_{1\lambda}(\eta)-q_{1\lambda}(\eta)\frac{\rmd}{\rmd\eta}p_{1\lambda}(\eta)=-\frac{2}{\eta^{2}-1}.
\end{equation}
Dividing across by $(p_{1\lambda})^{2}$ and integrating we obtain the following integral expression for $q_{1\lambda}(\eta)$
\begin{equation}
\label{eq:qint}
q_{1\lambda}(\eta)=2p_{1\lambda}(\eta)\int_{\eta}^{\infty}\frac{\rmd\xi}{(\xi^{2}-1)[p_{1\lambda}(\xi)]^{2}}\,\, ,
\end{equation}
where we have used the fact that $q_{1\lambda}(\eta)\rightarrow0$ as $\eta\rightarrow\infty$. One can easily verify that the series solution to $p_{1\lambda}(\xi)$ about $\xi=1$ is 
\begin{equation}
\label{eq:pseries}
p_{1\lambda}(\xi)=(\xi-1)^{1/2}+\case{1}{4}\lambda(\lambda+1)(\xi-1)^{3/2}+\Or(\xi-1)^{5/2}
\end{equation}
and therefore we have
\begin{equation}
\label{eq:pinvseries}
\frac{1}{[p_{1\lambda}(\xi)]^{2}}=\frac{1}{\xi-1}-\frac{l(l+1)}{2}+\Or(\xi-1).
\end{equation}
We now add and subtract the first two terms on the right-hand side above in the integral expression (\ref{eq:qint}),
\begin{eqnarray}
q_{1\lambda}(\eta)=&2p_{1\lambda}(\eta)\int_{\eta}^{\infty}\frac{\rmd\xi}{(\xi^{2}-1)}\Big(\frac{1}{[p_{1\lambda}(\xi)]^{2}}-\frac{1}{(\xi-1)}+\frac{\lambda(\lambda+1)}{2}\Big) \nonumber\\
&+2p_{1\lambda}(\eta)\int_{\eta}^{\infty}\frac{\rmd\xi}{(\xi^{2}-1)}\Big(\frac{1}{\xi-1}-\frac{\lambda(\lambda+1)}{2}\Big).
\end{eqnarray}
It is clear from Eqs.(\ref{eq:pseries}) and (\ref{eq:pinvseries}) that the first integral is convergent as $\eta\rightarrow1$, and the second integral can be evaluated explicitly. We thus obtain the following asymptotic expansion of $q_{1\lambda}(\eta)$ as $\eta\rightarrow1$,
\begin{eqnarray}
\label{eq:q1series1}
\fl q_{1\lambda}(\eta)=\frac{1}{(\eta-1)^{1/2}}+\case{1}{2}[1+\lambda(\lambda+1)](\eta-1)^{1/2}\ln(\eta-1) \nonumber\\
\fl \ \ +\Big\{2 I_{\lambda}+\case{1}{4}\lambda(\lambda+1)-\case{1}{2}[1+\lambda(\lambda+1)]\ln2\Big\}(\eta-1)^{1/2}+\Or((\eta-1)^{3/2}\ln(\eta-1))\,,
\end{eqnarray}
where
\begin{equation}
I_{\lambda}=\int_{1}^{\infty}\frac{\rmd \xi}{(\xi^{2}-1)}\Big(\frac{1}{[p_{1\lambda}(\xi)]^{2}}-\frac{1}{(\xi-1)}+\frac{\lambda(\lambda+1)}{2}\Big).
\end{equation}

Recall that we also had an alternate expression for $q_{1\lambda}(\eta)$ (\ref{eq:q1def}) which we expand about $\eta=1$ to get
\begin{eqnarray}
\label{eq:q1series2}
q_{1\lambda}(\eta)=\frac{1}{(\eta-1)^{1/2}}+\case{1}{2}[1+\lambda(\lambda+1)](\eta-1)^{1/2}\ln(\eta-1) \nonumber\\
+\Big\{[1+\lambda(\lambda+1)][\Psi(\lambda+1)+\gamma-\case{1}{2}\ln2-\case{1}{2}]+\case{1}{4}+\beta_{\lambda}\Big\}(\eta-1)^{1/2}\nonumber\\
+\Or((\eta-1)^{3/2}\ln(\eta-1))\, ,
\end{eqnarray}
where $\Psi(z)=\Gamma '(z)/\Gamma(z)$ and $\gamma$ is Euler's constant.

\begin{table}[htdp]
 \caption{The coefficients $\beta_{\lambda}$ for $\alpha=0.95$.}
\begin{center}
\begin{tabular}{| c || c | c | c | c | c |c|} \hline
  & $m$= 0   &    1   &    2     &     3    &     4    &    5   \\
\hline\hline
$l$=0                        & 0.119377314  &                          &                               &                                      &                                    & \\

\hline
$l$=1                       & 0.002779252  &  0.002318785   &                               &                                      &                                    & \\

\hline
$l$=2                        & 0.000144497  & 0.000127179    & 0.000112214      &                                      &                                    & \\

\hline
$l$=3                       & 0.000018348 &    0.000016749                     &     0.000015310      &  0.000014014            &                                    & \\

\hline
$l$=4                       & $0.000003975 $ &   $0.000003705$   &  $0.000003456$  &      $0.000003226$  &     $0.000003013$  & \\

\hline
$l$=5                       & $0.000001181 $ &     $0.000001115    $    &   $  0.000001053   $    &   $ 0.000000995 $    &  $0.000000941 $        & $0.000000890 $ \\

\hline

\end{tabular}
\end{center}
\label{default}
\end{table}

\begin{table}[htdp]
 \caption{The coefficients $\beta_{\lambda}$ for $\alpha=0.6$.}
\begin{center}
\begin{tabular}{| c || c | c | c | c | c |c|} \hline
  & $m$= 0   &    1   &    2     &     3    &     4    &    5   \\
\hline\hline
$l$=0                        & 0.119377314  &                          &                               &                                      &                                    & \\

\hline
$l$=1                       & 0.002779252  &  0.000344996   &                               &                                      &                                    & \\

\hline
$l$=2                        & 0.000144497  & 0.000033858    & 0.000010533     &                                      &                                    & \\

\hline
$l$=3                       & 0.000018348 &    0.000006342                     &     0.000002579      &  0.000001181           &                                    & \\

\hline
$l$=4                       & $0.000003975 $ &   $0.000001723$   &  $0.000000828$  &      $0.000000431$  &     $0.000000239$  & \\

\hline
$l$=5                       & $0.000001181 $ &     $0.000000592    $    &   $  0.000000319   $    &   $ 0.000000182 $    &  $0.000000109$        & $0.000000068 $ \\

\hline

\end{tabular}
\end{center}
\label{default}
\end{table}

Finally, comparing our two equivalent expressions for $q_{1\lambda}(\eta)$ given by (\ref{eq:q1series1}) and (\ref{eq:q1series2}) yields an expression for $\beta_{\lambda}$ that is numerically tractable:
\begin{equation}
\beta_{\lambda}=2 I_{\lambda}-[1+\lambda(\lambda+1)][\Psi(\lambda+1)+\gamma-\case{3}{4}]-\case{1}{2}.
\end{equation}

That the derivation of the $\beta_{\lambda}$ coefficients depends on the integral (\ref{eq:qint}) underlines the fact these coefficients are truly global in nature since the dependence on the boundary conditions is explicit here. If we considered the case where we place the black hole in a spherical box, then the upper limit of the integration above would be the radius of the box and we would obtain different $\beta_{\lambda}$ coefficients.

The integral $I_{\lambda}$ is most effectively calculated by splitting the range of integration into the regions $1\le\xi\le2$ and $2\le\xi<\infty$. In the former range, one can use a series solution to $p_{1\lambda}(\eta)$ in order to cancel the divergences in the integrand near $\eta=1$, and the integral can now be performed without difficulty. In the latter range, the first term cuts off exponentially as $\eta\rightarrow \infty$ since $p_{1\lambda}(\eta)\rightarrow\infty$ as $\eta\rightarrow\infty$. The remaining part of the integral can be performed explicitly. We have tabulated above some $\beta_{\lambda}$ coefficients for $\alpha=0.95$ and $\alpha=0.6$ for $l$ up to $l=5$. We note that for $m=0$, we have $\lambda = l$ an integer, and we retrieve the Candelas \cite{Candelas:1980zt} $\beta_{l}$ values.

\section*{Acknowledgements}
PT is supported by the Irish Research Council for Science, Engineering and Technology, funded by the National
Development Plan.

The authors would like to thank Cormac Breen for many helpful conversations. 

\section*{References}

\begin{thebibliography}{10}

\bibitem{Candelas:1980zt}
P.~Candelas,
\newblock {\em Phys.\ Rev.\ D}, 21:2185, 1980.

\bibitem{Candelas:1984pg}
P.~Candelas and K.W.~Howard,
\newblock {\em Phys.\ Rev.\ D}, 29:1618, 1984.

\bibitem{Anderson:1989vg}
P.R.~Anderson,
\newblock {\em Phys.\ Rev.\ D}, 39:3785, 1989.

\bibitem{JensenOttewill}
B.P.~Jensen and A.C.~Ottewill,
\newblock {\em Phys. Rev. D}, 39:1130, 1989.

\bibitem{JMO1}
J.G.~McLaughlin, B.P.~Jensen and A.C.~Ottewill,
\newblock {\em Phys. Rev. D}, 45:3002, 1992.

\bibitem{JMO2}
J.G.~McLaughlin, B.P.~Jensen and A.C.~Ottewill,
\newblock {\em Phys. Rev. D}, 51:5676, 1995.

\bibitem{Anderson:1990jh}
P.R.~Anderson,
\newblock {\em Phys.\ Rev.\ D}, 41:1152, 1990.

\bibitem{Winstanley:2007}
E.~Winstanley and P.M.~Young,
\newblock {\em Phys. Rev. D}, 77, 2008.

\bibitem{Christensen:1976vb}
S.M.~Christensen,
\newblock {\em Phys.\ Rev.\ D}, 14:2490, 1976.

\bibitem{DaviesSahni}
P.C.W.~Davies and V.~Sahni,
\newblock {\em Class.\ Quantum.\ Grav}, 5:1-17, 1988.

\bibitem{CSHorizon}
A.C.~Ottewill and P.~Taylor,
\newblock arXiv:1007:0051[gr-qc], to appear in {\em Phys.\ Rev.\ D}.


\bibitem{Erdelyi}
F.~Oberhettinger, A.~Erdelyi, W.~Magnus and F.G.~Tricomi,
\newblock {\em Higher Transcendental Functions}.
\newblock McGraw-Hill, New York, 1953.

\bibitem{Smith}
A.G.~Smith,
\newblock {\em Formation and Evolution of Cosmic Strings}.
\newblock eds G. W. Gibbons, S. W. Hawking and T. Vaschaspati, (Cambridge
  University Press), 1990.

\bibitem{Decanini:2008}
Y.~Decanini and A.~Folacci,
\newblock{\em Phys.\ Rev.\ D}, 78:044025, 2008.


\bibitem{BrownOttewill1986}
M.R.~Brown and A.C.~Ottewill,
\newblock {\em Phys.\ Rev.\ D}, 34:1776, 1986.


\bibitem{gradriz}
I.S.~Gradshteyn and I.M.~Ryzhik,
\newblock {\em Table of Integrals, Series and Products}.
\newblock Academic Press, 2000.

\bibitem{Weinstein}
A.~Weinstein,
\newblock {\em Amer. Math. Soc.}, 63:342, 1948.

\bibitem{Linet1977}
B.~Linet,
\newblock {\em Phys. Lett.}, 60A(5):395, 1977.

\bibitem{LinetCosmicString1987}
B.~Linet,
\newblock {\em Phys.Rev. D}, 35:536--539, 1987.

\end{thebibliography}
\bibliographystyle{unsrt}

\newpage

\end{document}